\documentclass[useAMS]{mn2e}

\usepackage{graphicx,color,soul}
\usepackage{latexsym}
\usepackage{multirow}
\usepackage{amsmath,amssymb}        
\usepackage[draft=false]{hyperref}

\title[Axisymmetric Bondi-Hoyle accretion onto a Schwarzschild Black Hole: shock cone vibrations]{Axisymmetric Bondi-Hoyle accretion onto a Schwarzschild Black Hole: shock cone vibrations}

\author[F. D. Lora-Clavijo and F. S. Guzm\'an]{F. D. Lora-Clavijo and F. S. Guzm\'an\thanks{E-mail:
fadulora@ifm.umich.mx; guzman@ifm.umich.mx} \\
	     Instituto de F\'{\i}sica y Matem\'{a}ticas, Universidad
              Michoacana de San Nicol\'as de Hidalgo.\\ Edificio C3, Cd.
              Universitaria, 58040 Morelia, Michoac\'{a}n,
              M\'{e}xico.}

\begin{document}


\date{\today}

\pagerange{\pageref{firstpage}--\pageref{lastpage}} \pubyear{2011}

\maketitle

\label{firstpage}


\begin{abstract}
We study numerically the  axisymmetric relativistic Bondi-Hoyle accretion of a supersonic ideal gas onto a fixed Schwarzschild background spaceÐtime described with horizon penetrating coordinates. We verify that a nearly stationary shock cone forms and that the properties of the shock cone are consistent with previous results in Newtonian gravity and former relativistic studies. The fact that the evolution of the gas is tracked on a spatial domain that contains a portion of the inner part of the black hole avoids the need to impose boundary conditions on a time-like boundary as done in the past. Thus, our approach contributes to the solution to the Bondi-Hoyle accretion problem at the length scale of the accretor in the sense that the gas is genuinely entering the black hole. As an astrophysical application, we study for a set of particular physical parameters, the spectrum of the shock cone vibrations and their potential association with QPO sources.
\end{abstract}




\section{Introduction}

Bondi-Hoyle accretion is the process of evolution of an infinite uniform wind of gas moving near a central object, or conversely a compact object moving on a uniform gas with constant velocity \cite{Bondi}. This process is associated to various phenomena related to gas processes near stars or compact objects. It shows interesting properties when considered within the Newtonian and relativistic regimes, which have been explored based on several numerical studies in the classical regime ruled by Newtonian gravity  \cite{Shima,Matsuda1,Fryxell,Sawada,Matsuda2,Matsuda3,Ruffert1,Ruffert2,Ruffert3,Ruffert4,Ruffert5,Benensohn,Nagae,Blondin}, in which the most important subject are the consequences on the morphology of the wind and the supersonic shocks that develop; a summary of results under Newtonian gravity can be found in \cite{Foglizzo}.

Unlike in the Newtonian regime, the relativistic approach allows the study of the Bondi-Hoyle accretion in regions where the gravitational field is strong, for instance, near the event horizon of a black hole. Some studies have been pushed forward in this direction. The first one was performed by  \cite{Petrich}; in this work they carry out axially symmetric numerical simulations in order to study the different accretion patterns  developed by the gas during the accretion onto a black hole. Later on,  \cite{Fontaxi,Font1,Font2,Font3} revisited the results using high resolution shock capturing schemes that have shown to be much more accurate methods than those used in the past consisting in the addition of artificial viscosity. It is worth pointing out that in all these papers the morphology of subsonic and supersonics flows was the most important subject, because when the wind has supersonic velocity a high density shock cone forms on the opposite side of the wind source, and such cone has interesting properties like resonant oscillations or flip-flop motion. On the other hand, in the astrophysical context, a recent work by \cite{Donmez} dedicates important efforts to study a potential relation between QPOs and  the behavior of the gas density in the relativistic Bondi-Hoyle accretion. 

Considering full general relativity, Bondi-Hoyle accretion has been studied by \cite{Farris} in the context of the merger of supermassive black hole binaries. Also axisymmetric  magnetohydrodynamics Bondi-Hoyle accretion was studied in \cite{penner1}. On the other hand \cite{Zanotti} studied the case of the relativistic Bondi-Holye accretion considering radiative processes and the case of  ultrarelativistic axisymmetric Bondi-Hoyle accretion was recently considered in \cite{penner2}.

In this paper we present a relativistic study of supersonic axisymmetric Bondi-Hoyle accretion onto a spherically symmetric black hole described with horizon penetrating coordinates. We assume the condition that the gas is a test fluid and does not distort the geometry of the space-time. Previous studies on the subject involve the use of coordinates that are singular at the event horizon, and therefore the evolution problem has to be set on a domain that requires an artificial inner boundary outside the black hole's horizon, which requires the implementation of efficient boundary conditions there, specially when such boundary is close to the event horizon where the state variables tend to diverge, in order to prevent errors to propagate toward the numerical domain and consequently affect the numerical calculations. What the use penetrating coordinates offers, is the possibility to define the inner boundary inside the black hole's horizon and furthermore to avoid the implementation of boundary conditions there, because the light cones using penetrating coordinates remain open and material particles move naturally toward inside the black hole. In other words, the fluid is allowed to fall into the black hole. Some basic models of spherical accretion of dark matter by supermassive black holes already use this condition \cite{GuzmanLora1,GuzmanLora2}.

Even though our approach represents an improvement to the complete study of Bondi-Hoyle accretion, it is only an alternative to solve the problem at the inner boundary related to the length scale of the accretor. At a different scale, the accretion radius is defined to approximately decide when a particle falls into the compact object or not, and such scale is determined by the velocity of the wind and the equation of state of the gas. Traditionally, the Bondi-Hoyle problem is attended in this second scale, by assuming the accretor is point-like, whereas in our case we are exactly doing the opposite. Numerically it seems that one has to choose between these two regimes, that is, it is possible to apply wind accretion to astrophysical processes if realistic wind velocities are considered, and consequently the accretor length scale is small enough as to be considered a point-like source that cannot be numerically resolved; conversely, if one wants to resolve the accretor (like in our case) the accretion radius length scale is unresolved unless the velocity of the wind is assumed to be high. This difficulty of studying both scales at the same time is the main reason why the relativistic wind accretion has not been fully resolved at the moment.

The association of numerical results on Bondi-Hoyle accretion onto black holes to astrophysical observations involves a series of parameters that in principle should be astrophysically motivated.  
For instance the adiabatic index and internal energy of the gas, and the velocity of the wind if one wants to infer the mass and angular momentum of the black hole, or viceversa; furthermore, properties of the gas associated to heat transfer of cooling processes expected to happen would involve an extra set of parameters that should be observationally motivated. Perhaps the most solid bounds among all these parameters is the relative velocity between the accretor and the wind, which is of the order of 100-1000 km/s in binaries and may achieve from a few thousands of km/s \cite{Gonzalez} in superkicked black holes resulting from the collision of two black holes with collimated emission of gravitational radiation up to very recent results indicating 15000 km/s  \cite{kick} in similar scenarios.
 
The paper is organized as follows. In section \ref{sec:GRHD} we define the relativistic Euler equations describing the fluid onto a fixed Schwarzschild black hole in Eddington-Finkelstein coordinates, whereas in section \ref{sec:numerics} we explain the numerical methods we use to solve the relativitic Euler equations on a curved background space-time; in section \ref{sec:Morpho} we show the morphology and mass accretion rates for the accretion of an axisymmetric homogenous wind and in \ref{sec:QPOs_results} we analyze the frequencies associated to the rest mass density oscillations inside the shock cone. Finally in \ref{sec:conclusions} we present our conclusions.


\section{Relativistic Hydrodynamic Equations}
\label{sec:GRHD}
 
For a generic space-time, relativistic Euler equations can be derived from the local conservation of the rest mass and the stress-energy tensor $T_{\mu \nu}$ of the fluid model considered:

\begin{eqnarray}
\nonumber && \nabla_{\nu} (T^{\mu \nu}) = 0, \\
&& \nabla_{\nu} (\rho_{0}u^{\nu}) = 0, \label{eq:cons}
\end{eqnarray}

\noindent where $\rho_{0}$ is the rest mass density, $u^{\mu}$ is the four-velocity of the fluid and $\nabla_{\nu}$ is the covariant derivative consistent with the four-metric $g_{\mu \nu}$ of the space-time \cite{Misner}. Notice that our calculations assume geometric units where $G=c=1$. We consider the gas  accreted by the black hole is a perfect fluid, which can be described by the following stress-energy tensor 

\begin{equation} \label{eq:SET}
T_{\mu \nu} = \rho_{0}hu^{\mu} u^{\nu} + pg^{\mu \nu},
\end{equation}

\noindent where $p$ is the pressure and $h$ the relativistic specific enthalpy given by

\begin{equation} \label{eq:enthalpy}
h = 1 + \epsilon + \frac{p}{\rho_{0}}.
\end{equation}

\noindent where $\epsilon$ is the rest frame specific internal energy density of the fluid.

In order to track the evolution of the fluid it is convenient to write down the relativistic Euler equations as flux balance laws \cite{Marti,Banyuls,Font}, for which we require the space-time metric  to be written in  the 3+1 splitting approach of general relativity. In this paper we consider the accretion onto a spherically symmetric black hole for which the most general line element reads

\begin{eqnarray}
ds^2 &=& -(\alpha^2 - \gamma_{rr}\beta^r \beta^r)dt^2 + 2\gamma_{rr}\beta^r dr dt + \gamma_{rr}dr^2 \nonumber\\
&+& \gamma_{\theta\theta} (d\theta^2 + \sin^2 \theta d\phi^2), \label{eq:lineelement}
\end{eqnarray}

\noindent where $\alpha$ is the lapse function, the shift vector has only one component $\beta^i=(\beta^r,0,0)$, $\gamma_{ij}$ are the components of the spatial metric associated with the space-like hypersurfaces and  $x^\mu=(t,r,\theta,\phi)$ are coordinates of the space-time that we choose to be spherical in the spatial part. 

A fingerprint of our paper is that we use penetrating coordinates, because the previous research uses mostly Schwarzschild coordinates, which are singular at the event horizon, and most authors implement an artificial time-like boundary near the event horizon in a region where the gravitational field is extremely strong and numerical boundary conditions in such region might not be as appropriate as in the far region. We instead use horizon penetrating Eddington-Finkelstein slices to describe the black hole, for which $\alpha$, $\beta^i$ and $\gamma_{ij}$ read

\begin{eqnarray} 
\nonumber \alpha&=&\frac{1}{\sqrt{1+\frac{2M}{r}}}, \\ 
\nonumber \beta^i&=&\left[ \frac{2M}{r}\left( \frac{1}{1+\frac{2M}{r}}\right),0,0\right], \\
\gamma_{ij}&=&diag[1+\frac{2M}{r},r^2,r^2\sin^2\theta].
\end{eqnarray}

\noindent Since we consider the test field approximation to be valid in our numerical experiments, there is no need to evolve the geometry of the space-time, instead we keep our background space-time fixed and describe the evolution of the fluid on top of such space-time. The general equations obtained from (\ref{eq:cons}) for a fluid with axisymmetry in spherical coordinates for metric (\ref{eq:lineelement}) read

\begin{equation}
\partial_{t}  {\bf U} + \partial_{r} (\alpha {\bf F}^{r}) + \partial_{\theta} (\alpha {\bf F}^{\theta}) = {\alpha \bf S} - \frac{\partial_r \sqrt{\gamma}}{\sqrt{\gamma}}(\alpha {\bf F}^{r}) - \frac{\partial_\theta \sqrt{\gamma}}{\sqrt{\gamma}}(\alpha {\bf F}^{\theta}). \label{eq:HydroEvolve}
\end{equation} 
 
\noindent Here ${\bf U}$ is a vector of conservative variables, ${\bf F}^i$ are the fluxes in the spatial directions and ${\bf S}$ is a vector of sources. These quantities are defined in terms of both the primitive variables $(\rho_{0},v^{r},v^{\theta},\epsilon)$ and the conservative variables themselves as follows:

\begin{eqnarray}
&& {\bf U} = \left [
\begin{array}{c}
D \\
J_r \\
J_\theta \\
\tau 
\end{array}
\right] =
\left [
\begin{array}{c}
\rho_0 W \\
\rho_0 h W^2 v_r \\
\rho_0 h W^2 v_\theta \\ 
\rho_0 h W^2 -p - \rho_0 W
\end{array}
\right] , \label{eq:qvector} \\ \nonumber
\\ 
&&{\bf F}^ r= 
\left [
\begin{array}{c}
\left (v^r - \frac{\beta^r}{\alpha} \right )D \\
\left (v^r - \frac{\beta^r}{\alpha} \right )J_r + p \\
\left (v^r - \frac{\beta^r}{\alpha} \right )J_\theta \\
\left (v^r - \frac{\beta^r}{\alpha} \right )\tau + v^r p 
\end{array}
\right] , \label{eq:fvector} \\ \nonumber
\\
&&{\bf F}^ \theta= 
\left [
\begin{array}{c}
v^\theta D \\
v^\theta J_r \\
v^\theta J_\theta + p\\
v^\theta( \tau + p) 
\end{array}
\right] , \label{eq:fvector2} \\ \nonumber
\\
&&{\bf S} = 
\left [
\begin{array}{c}
0\\
T^{\mu \nu} g_{\nu \sigma} \Gamma^{\sigma}_{\mu r} \\
T^{\mu \nu} g_{\nu \sigma} \Gamma^{\sigma}_{\mu \theta} \\
T^{\mu 0} \partial_{\mu}\alpha - \alpha T^{\mu \nu} \Gamma^{0}_{\mu \nu}
\end{array}
\right]. \label{eq:svector}
\end{eqnarray}  

\noindent In these expressions,  $\gamma=det(\gamma_{ij})$ is the determinant of the spatial 3-metric, $\Gamma^{\sigma}{}_{\mu \nu}$ are the Christoffel symbols of the space-time and  $v^{i}=(v^r,v^\theta,0)$ is the 3-velocity measured  by an Eulerian observer and defined in terms of the spatial part of the 4-velocity $u^{i}$, as $v^{i}=\frac{u^{i}}{W} + \frac{\beta^{i}}{\alpha}$, where $W$ is the Lorentz factor $W=\frac{1}{\sqrt{1-\gamma_{ij} v^{i}v^{j}}}$.
  
The system of equations (\ref{eq:HydroEvolve}-\ref{eq:svector}) is a set of four equations for either the primitive variables $\rho_0,v^r, v^{\theta},\epsilon,p$ or the conservative variables $D,J_r,J_{\theta},\tau$ and $p$ or $\epsilon$. Therefore it is necessary to close the system of equations. As usual we close the system using an equation of state relating $\epsilon=\epsilon(\rho_{0},p)$. In this work, we choose the gas to obey an ideal gas equation of state given by

\begin{equation}
\epsilon = \frac{p}{(\Gamma-1)\rho_{0}}, \label{eq:EOS}
\end{equation}

\noindent where $\Gamma$ is the ratio of specific heats or adiabatic index. The relativistic speed of sound $c_{s}$ plays an important role in the definition of the initial conditions of the wind of gas we evolve, and for the equation of state (\ref{eq:EOS}) can be written as 

\begin{equation}
c_{s}^{2} = \frac{p\Gamma(\Gamma-1)}{p\Gamma + \rho_{0}(\Gamma -1)}, \label{eq:SoundSpeed}
\end{equation}

\noindent where its asymptotic value or its maximum permitted value is $c_{s_{max}}=\sqrt{\Gamma -1}$. 

\section{Initial data and numerical methods}
\label{sec:numerics}

\subsection{Evolution}

We evolve the system of equations (\ref{eq:HydroEvolve}-\ref{eq:svector}) using a high resolution shock capturing method \cite{LeVeque}. We specially use the HLLE Riemann solver \cite{HLLE} in combination with the minmod linear piecewise reconstructor. For the integration in time we used a second order Runge-Kutta integrator.

We evolve the gas on the domain $([r_{exc},r_{max}]\times[0,\pi])\times [0,2\pi)$ in spherical coordinates using a two dimensional code, and a uniformly spaced grid along coordinates  $r$ and $\theta$. The radial domain deserves a careful description. On the one hand, $r_{exc}$ defines a spherical boundary that we can choose inside the black hole's event horizon called an excision boundary originally implemented to study the evolution of black holes \cite{SeidelSuen1992} and recently incorporated to the study of hydrodynamics \cite{Hawke}. That is, a chunk of the domain inside the black holeÕs horizon is removed from the numerical domain from $0 < r \leq r_{exc}$ in order to avoid the black hole singularity and the steep gradients of metric functions near there; since the light cones inside the horizon (located at $r=2M$) point towards the singularity, there is no need to impose boundary conditions at $r = r_{exc}$, instead, the fluid simply gets off the domain through such boundary. We choose the excision radius to be $r_{exc} = 1.5M$,  which is both far enough from the singularity at $r = 0$ and provides a buffer zone $1.5M < r < 2M$ for the fluid inside the horizon to flow smoothly from the horizon towards the excision boundary. This is a rather improvement compared to previous results where Schwarzschild coordinates are used and the excision boundary is chosen outside the black hole, in a time-like region that requires the implementation of boundary conditions.

On the other hand $r_{max}$ defines an exterior spherical boundary. We distinguish between two hemispheres called downstream and upstream as in \cite{Matsuda1,Fontaxi}. The upstream hemisphere is defined as the part of the sphere where the gas is entering the domain and the downstream is the hemisphere through which the gas leaves the domain. In the downstream hemisphere we impose outflow boundary conditions and in the upstream boundary we always consider the initial asymptotic value of all variables as set initially. 

Another important issue is that $r_{max}$ has to be bigger than the accretion radius $r_{acc}$ (see below) if one wants  shocks to form, and experience tells us that using $r_{max} \sim 10r_{acc}$ provides adequate time and space scales to see the shock cone until a nearly stationary stage.

At the axis defined by $\theta=0,\pi$ the sources are extrapolated from the cell next to the boundary and demand $v^{\theta}$ to be odd there. We also implemented an atmosphere, that is, we define a minimum value of $\rho_0$ that avoids the divergence of the specific enthalpy and errors that propagate to the other variables. We set the density of the atmosphere to $10^{-11}$, which we found to allow accuracy and consistency of our numerical results. Even though in our numerical experiments we do not measure such a small rest mass density during the stationary regime, the implementation of the atmosphere works only to prevent the density to be small in rarefaction zones that may form before the stationary stage. Finally, since the fluxes depend both on the primitive and the conservative variables (see (\ref{eq:fvector}) and (\ref{eq:fvector2})) we reconstruct the pressure in terms of the conservative and the other primitive variables using a Newton-Raphson algorithm.

We performed tests of our code for different radial domains and resolutions and found optimal number of cells depending on the different models, being particularly interested in maintaining a sufficiently large domain preserving the spatial resolution within a convergence regime. Also in the appendix we present a set of more numerical standard tests for our numerical implementation.

\subsection{Initial data}
 
We set the wind to move along the direction of $z$ initially, with constant density and pressure. We characterize the initial velocity field $v^i$ in terms of the asymptotic initial velocity $v_{\infty}$ as done in \cite{Fontaxi}:
 
 \begin{eqnarray}
 v_r &=& \frac{1}{\sqrt{\gamma^{rr}}} v_{\infty} \cos \theta, \\ \label{eq:vr}
 v_\theta &=& -\frac{1}{\sqrt{\gamma^{\theta \theta}}} v_{\infty} \sin \theta, \label{eq:vt}
 \end{eqnarray}
 
 \noindent where the relation $v^2=v_iv^i=v_{\infty}^{2}$ is satisfied. With this conditions, the gas initially with constant rest mass density moves along the $z$-direction filling the numerical domain. 
 
The initial density and pressure profiles are chosen in such a way that relation (\ref{eq:SoundSpeed}) is satisfied. In order to do this, we fix the initial density to a constant $\rho_{ini}$ and give an asymptotic value for the sound speed $c_{s_{\infty}}$, then using equation (\ref{eq:SoundSpeed}) the initial pressure can be obtained as follows
 
 \begin{equation}
 p_{ini} = \frac{c^2_{s_{\infty}} \rho_{ini} (\Gamma -1)}{\left( \Gamma (\Gamma -1) - c^2_{s_{\infty}} \Gamma \right)}. \label{eq:inipress}
 \end{equation}  
 
 \noindent A subtle point here is that we choose a value for $c_{s_{\infty}}$ to construct the initial pressure, with the only condition that in order to avoid negative or zero pressures we need the condition $c_{s_{\infty}}<\sqrt{\Gamma -1}$ to hold. Finally, using the equation of state (\ref{eq:EOS}) we calculate the initial internal energy $\epsilon_{ini}$.

At this point $v_{\infty}$ and $c_{s}{}_{\infty}$ are two important parameters to set up the initial velocity field, and we find useful to define the relativistic Mach number at infinity in order to parametrize our initial data ${\cal M}^{R}_{\infty}=W v_{\infty} / W_s c_{s}{}_{\infty} = W {\cal M}_{\infty}/W_s$, where $W$ is the Lorentz factor of the gas $W_s$ is the Lorentz factor calculated with the speed of sound and ${\cal M}_{\infty}$ is the asymptotic Newtonian Mach number we use to parametrize our initial configurations. When this number is bigger than one it is said that the flow is supersonic an otherwise subsonic. In the supersonic case a high density shock cone forms and because the density in the cone oscillates the matter transport of such oscillations could be associated to QPO sources.

A very important scale is the accretion radius defined by
  
 \begin{equation}
 r_{acc}=\frac{M}{v^2_\infty + c^2_{s_{\infty}} }, \label{eq:racc}
 \end{equation}
 
\noindent  otherwise all the gas in the domain would be accreted and such domain would then be inadequate to observe a shock cone.  This radius is the Bondi accretion radius, where $M$ is the accretor mass, in our case a black hole of mass $M=1$ \cite{Petrich}.

The parameter space to study the Bondi-Hoyle accretion is enormous and the free parameters are the initial rest mass density of the wind $\rho_{ini}$, the value of $\Gamma$, $v_{\infty}$, $c_{s}{}_{\infty}$ or equivalently the internal energy $\epsilon_{ini}$ of the gas or ${\cal M}$. We decide to fix the initial rest mass density $\rho_{ini}=10^{-6}$ in units where the mass of the black hole is $M=1$, which works in the test fluid approach we are assuming, although there are other precedents where a much higher density is used \cite{Donmez}. We also use this density because for a range between $10^{-8} < \rho_{ini} <10^{-4}$ we find the shock cone and its oscillatory behavior. In Table \ref{tab:InitialData} we present the parameter space we explore. Another possible parameter would be the mass of the black hole, nevertheless we choose units in which $M=1$, and automatically spatial and time measures will be in units of $M$; then our results are mass independent, and in order to specialize to a particular astrophysical case where $M$ is given in units of solar masses, we only need to rescale the units of space and time appropriately. 
We stress that we use parameters that allow the numerical track of the accretion process, which is restricted mainly by the size of the numerical domain; explicitly, if one assumes that either the gas or the black hole moves with speed of the order of $v_{\infty} \sim > c_{s\infty}\sim 100 km/s$, then $r_{acc}/M \sim 10^6$, then assuming the outer boundary of the numerical domain is at $r_{max} \sim 10r_{acc}$ it would be $\sim 10^7$ times the size of the black hole radius. In the most optimistic case, when a black hole is moving with a velocity of the order of $10^4km/s$ as a result of a superkick as response to the collimated emission of gravitational radiation \cite{kick}, the accretion radius would be of the order of $r_{acc}\sim 100M$ and $r_{max} \sim 1000M$. This is at the moment a restriction to carry out accurate numerical calculations, because on the one hand one needs a considerably big domain and on the other one needs enough resolution as to resolve the black hole. What we do here is that in order to cover a numerical domain using the necessary resolution to achieve convergence, we require the domain to be smaller than for the aforementioned astrophysical scenarios. We do this by considering rather high velocities of the gas that imply smaller values of $r_{acc} \sim 10M$ (see (\ref{eq:racc})).

\begin{table}
\caption{\label{tab:InitialData} Set of parameters we use for our numerical experiments. We use different values of $c_{s_{\infty}}$ and the same density $\rho_{ini} = 10^{-6}$ in all cases. The model in bold face was not solved numerically due to the computer  limitations described in the text, instead it is an extrapolation with the other similar cases to Mach 1. We use the two values of $\Gamma$ to test the shock formation and accretion, and use only $\Gamma=4/3$ for the QPO model.}
\begin{tabular}{ccccc} \hline
\hline
\\
Model & $\Gamma$ & $v_\infty $ & ${\cal M}_{\infty}$ & $r_{acc}$   \\ 
\\
\hline
\hline
&&$c_{s_{\infty}}=0.1$&&
\\
\hline
\hline
$M_{1a}$    &  $     4/3    $  & $    0.5    $ & $       5        $  & $3.84615$  
\\
\hline
$M_{1b}$    &  $     5/3    $  & $    0.5    $ & $       5        $  & $3.84615$  
\\
\hline
$M_{2a}$    &  $     4/3    $  & $    0.4    $ & $       4        $  & $5.88235$   
\\
\hline
$M_{2b}$    &  $     5/3    $  & $    0.4    $ & $       4        $  & $5.88235$  
\\
\hline
$M_{3a}$    &  $     4/3    $  & $    0.3    $ & $       3        $   & $10$  
\\
\hline
$M_{3b}$    &  $     5/3    $  & $    0.3    $ & $       3        $  & $10$ 
\\
\hline
$M_{4a}$    &  $     4/3    $  & $    0.2    $ & $       2        $  &  $20$ 
\\
\hline
$M_{4b}$    &  $     5/3    $  & $    0.2    $ & $       2        $  & $20$  
\\
\hline
$M_{5}$    &  $    4/3    $  & $    0.1    $ & $      1        $  &  $50$  
\\
\hline
\hline
& &$c_{s_{\infty}}=0.08$ & &  
\\
\hline
\hline
$M_{6}$    &  $     4/3    $  & $    0.4    $ & $       5        $  &   $6.00926$ 
\\
\hline
$M_{7}$    &  $     4/3    $  & $    0.32    $ & $       4        $  &   $9.19118$  
\\
\hline
$M_{8}$    &  $     4/3    $  & $    0.24    $ & $       3        $   &  $15.625$
\\
\hline
$M_{9}$    &  $     4/3    $  & $    0.16    $ & $       2        $  &  $31.25$   
\\
\hline
$M_{10} $    &  $   4/3    $  & $  0.08    $ & $      1   $  &  $78.125 $   
\\
\hline
\hline
& &$c_{s_{\infty}}=0.05$ & &  
\\
\hline
\hline
$M_{11}$    &  $     4/3    $  & $    0.25    $ & $       5        $  &    $15.3846$
\\
\hline
$M_{12}$    &  $     4/3    $  & $    0.2    $ & $       4        $  &    $23.5294$ 
\\
\hline
$M_{13}$    &  $     4/3    $  & $    0.15    $ & $       3        $   &  $40$
\\
\hline
$M_{14}$    &  $     4/3    $  & $    0.1    $ & $       2        $  &   $80$  
\\
\hline
${\bf M_{15} }$    &  $ {\bf    4/3 }   $  & $   {\bf 0.05 }   $ & $   {\bf    1 }       $  &   ${\bf 200}$  
\\
\hline
\hline
\end{tabular}
\end{table}

\subsection{Diagnostics}

In order to diagnose the physical quantities of the system in our simulations, we implement detectors located at various radii in the numerical domain, that is, we define spheres where we calculate scalar quantities. In particular, we track the mass accretion rate as a function of time. In order to do so we calculate the mass accretion rate on a sphere specialized to the case of a spherically symmetric space-time in spherical coordinates

\begin{equation}
\Dot{M}_{acc} = -2\pi \int D r^2 \left( v^r - \frac{\beta^r}{\alpha} \right) \sin \theta d \theta, \label{eq:MAccRate}
\end{equation}

\noindent at various spherical surfaces, including the black hole event horizon. The mass accretion rate helps diagnosing when the accretion has stabilized.

The most important quantity we measure is the rest mass density of the gas along the axis inside the shock cone, that is at $\theta=0$ and at the location of the different detectors. This scalar is used to measure how the density oscillates inside the shock cone. At post-step we perform a Fourier transform of this scalar in order to study the predominant oscillation modes of the density and study the relation between such oscillation and its potential relation to QPO sources.

\section{Properties of the shock cone and mass accretion rates}
\label{sec:Morpho}

Depending on the properties of the gas flow, a shock cone appears when the wind is supersonic. This shock cone is a region where the density is significantly higher than the density of the wind itself and it forms behind the black hole in the opposite side of the source of the wind.

\begin{figure*}
\includegraphics[width=4cm]{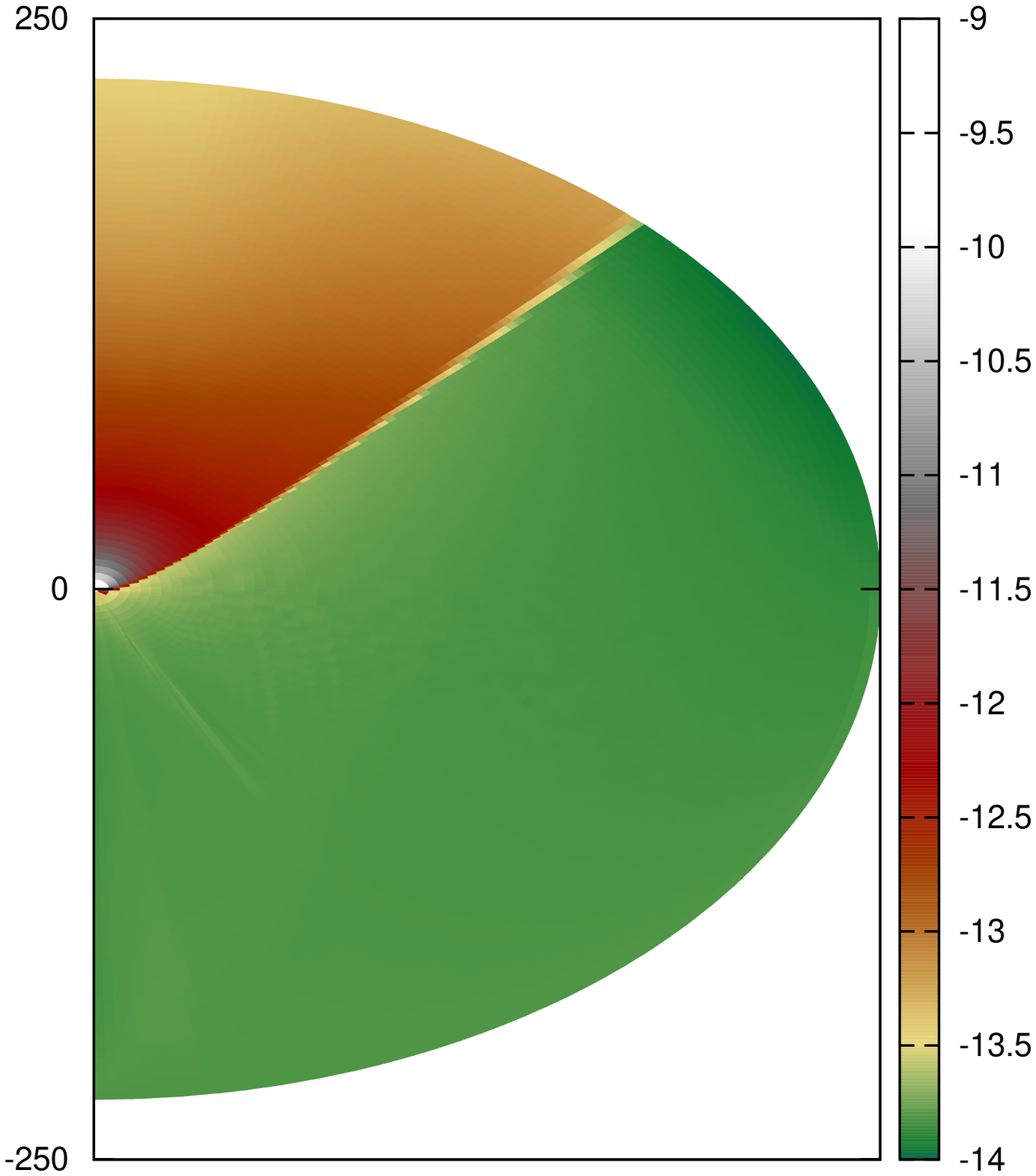}
\includegraphics[width=4cm]{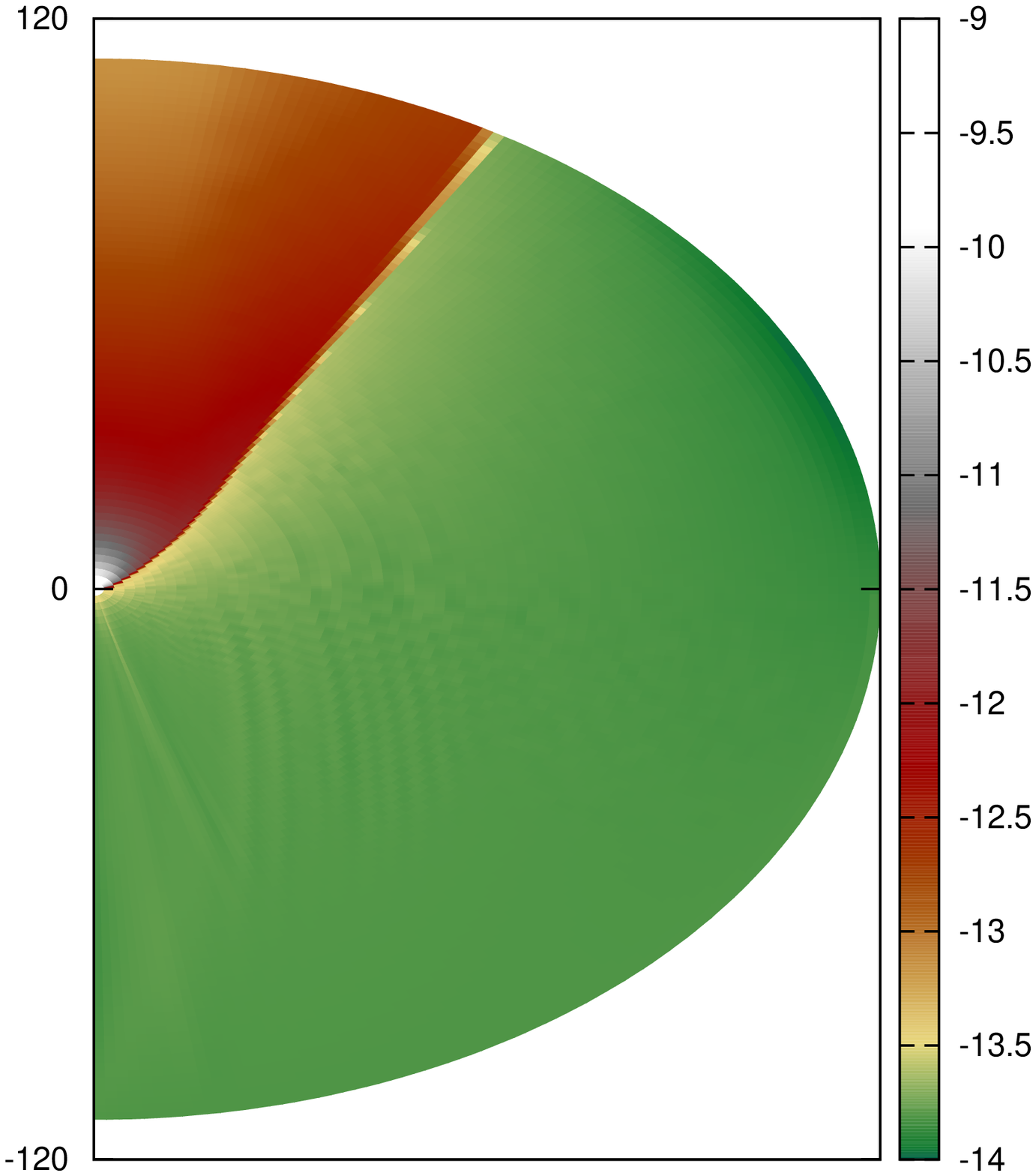}
\includegraphics[width=4cm]{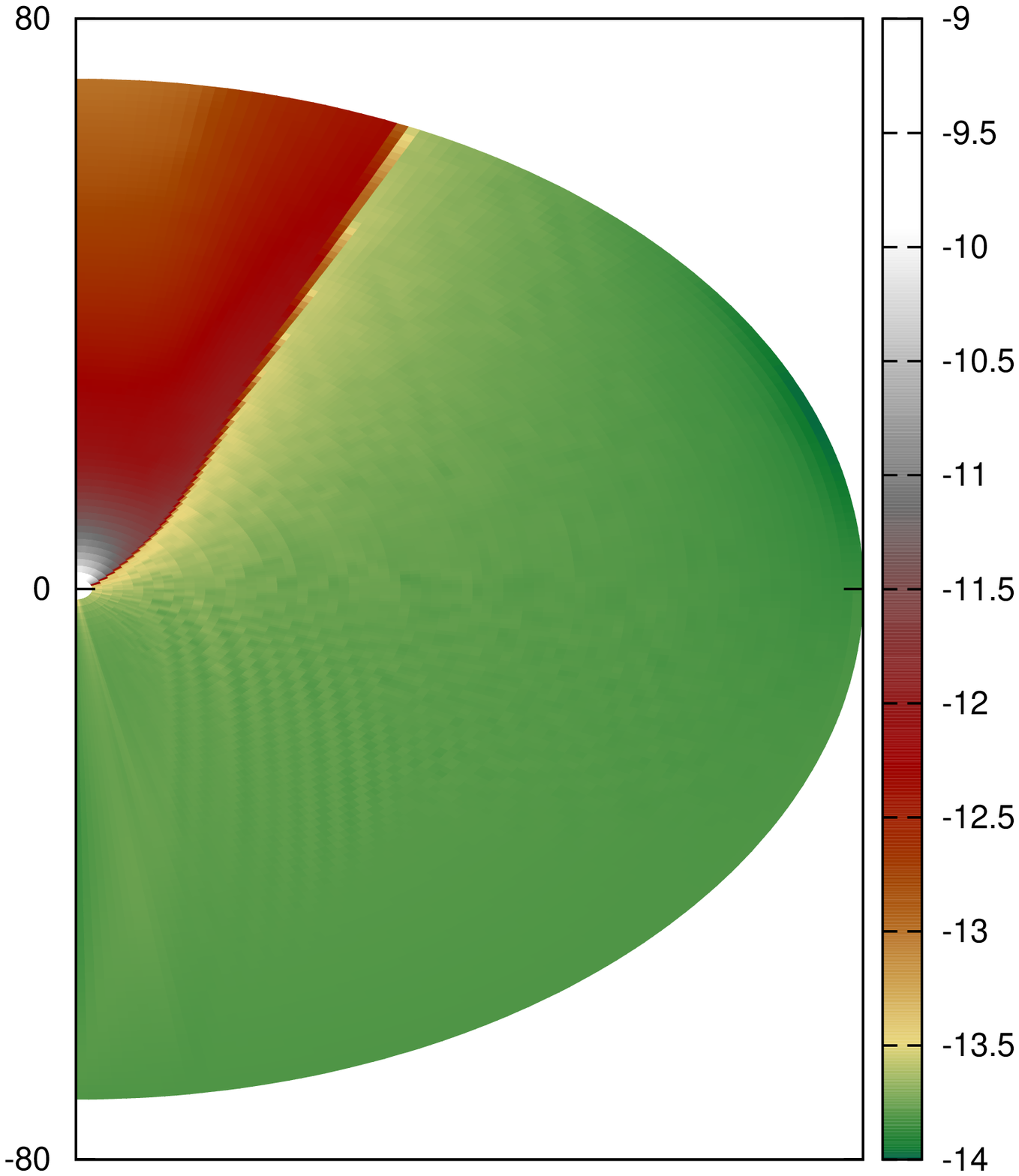}
\includegraphics[width=4cm]{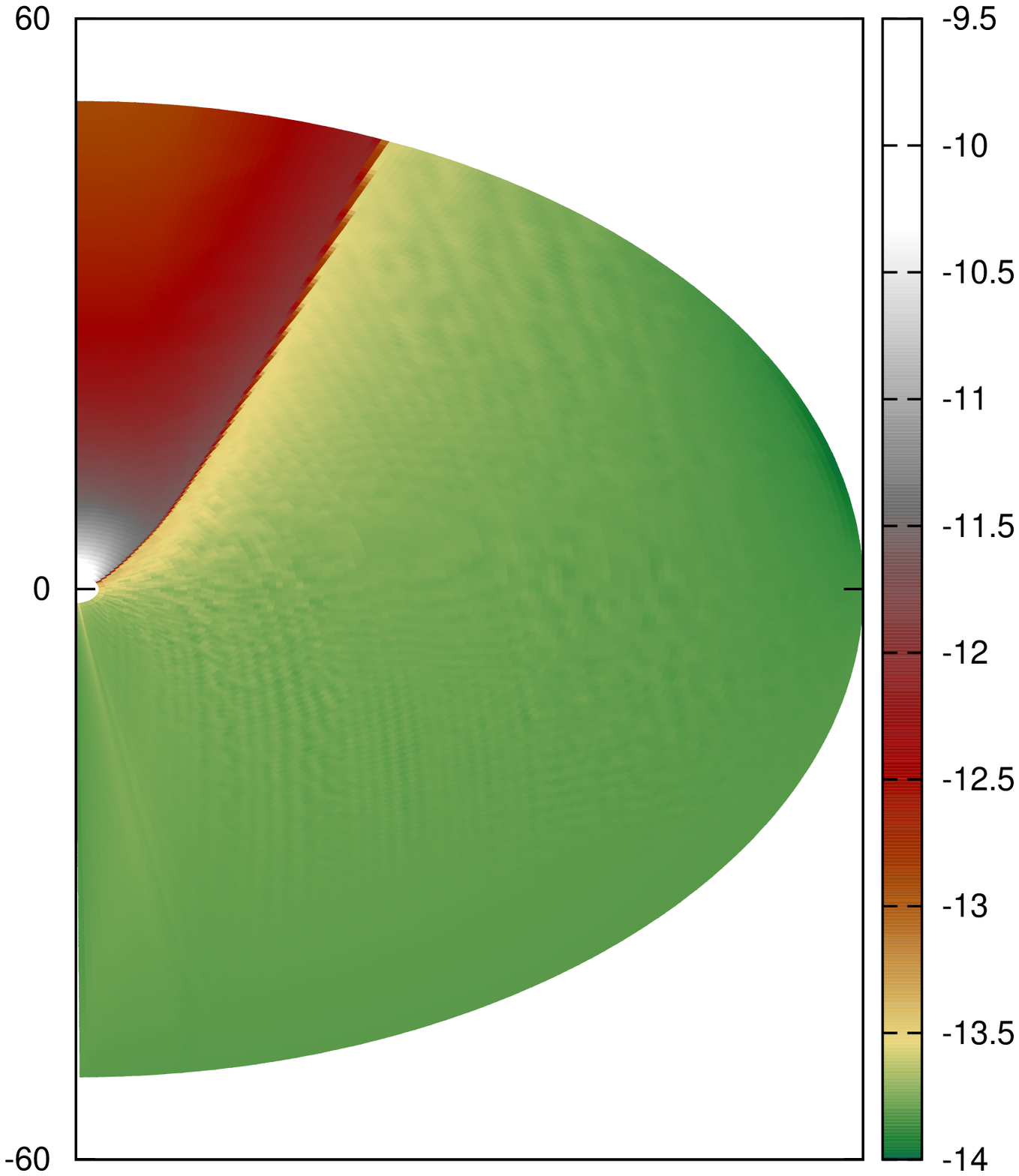}
\includegraphics[width=4cm]{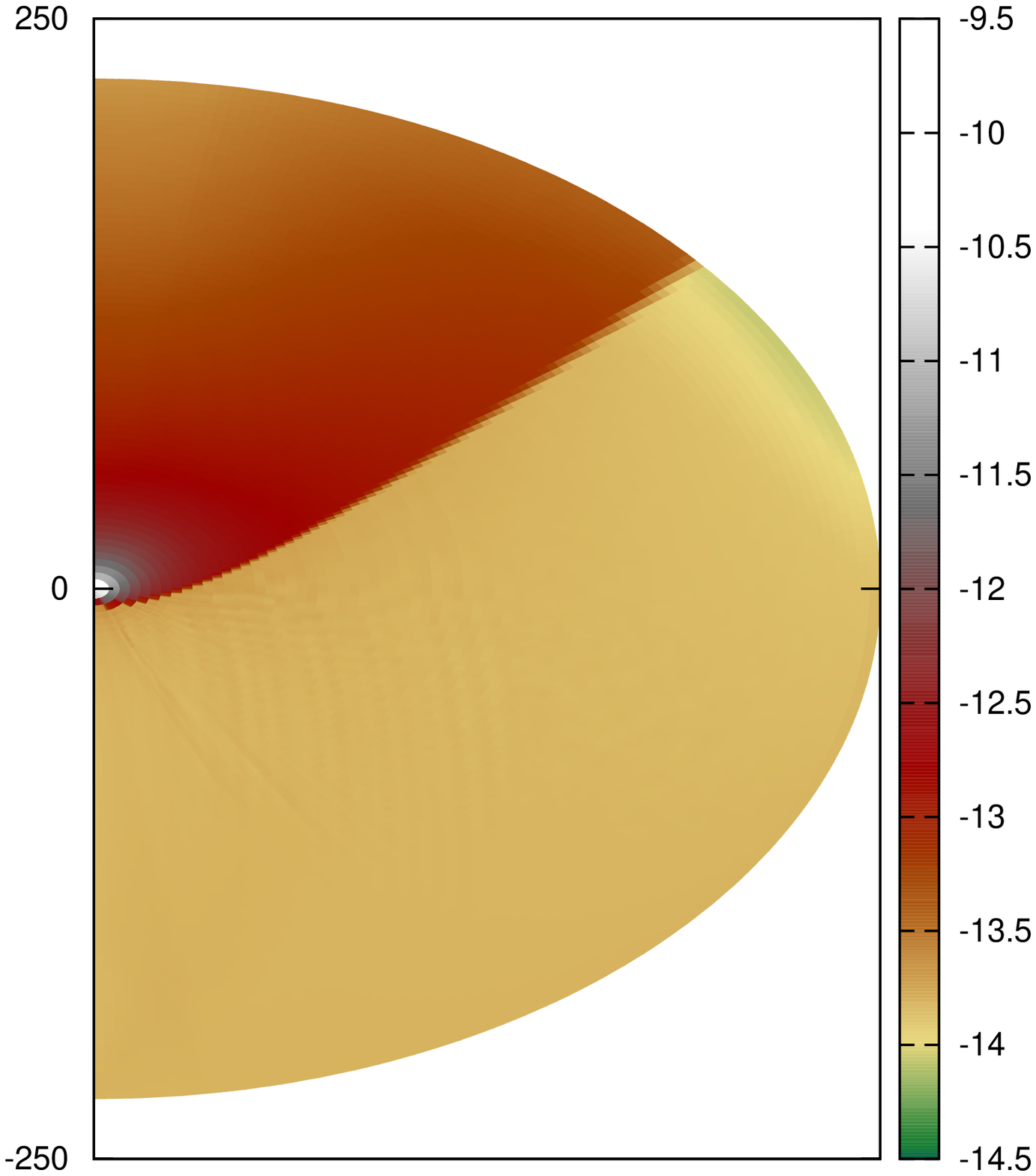}
\includegraphics[width=4cm]{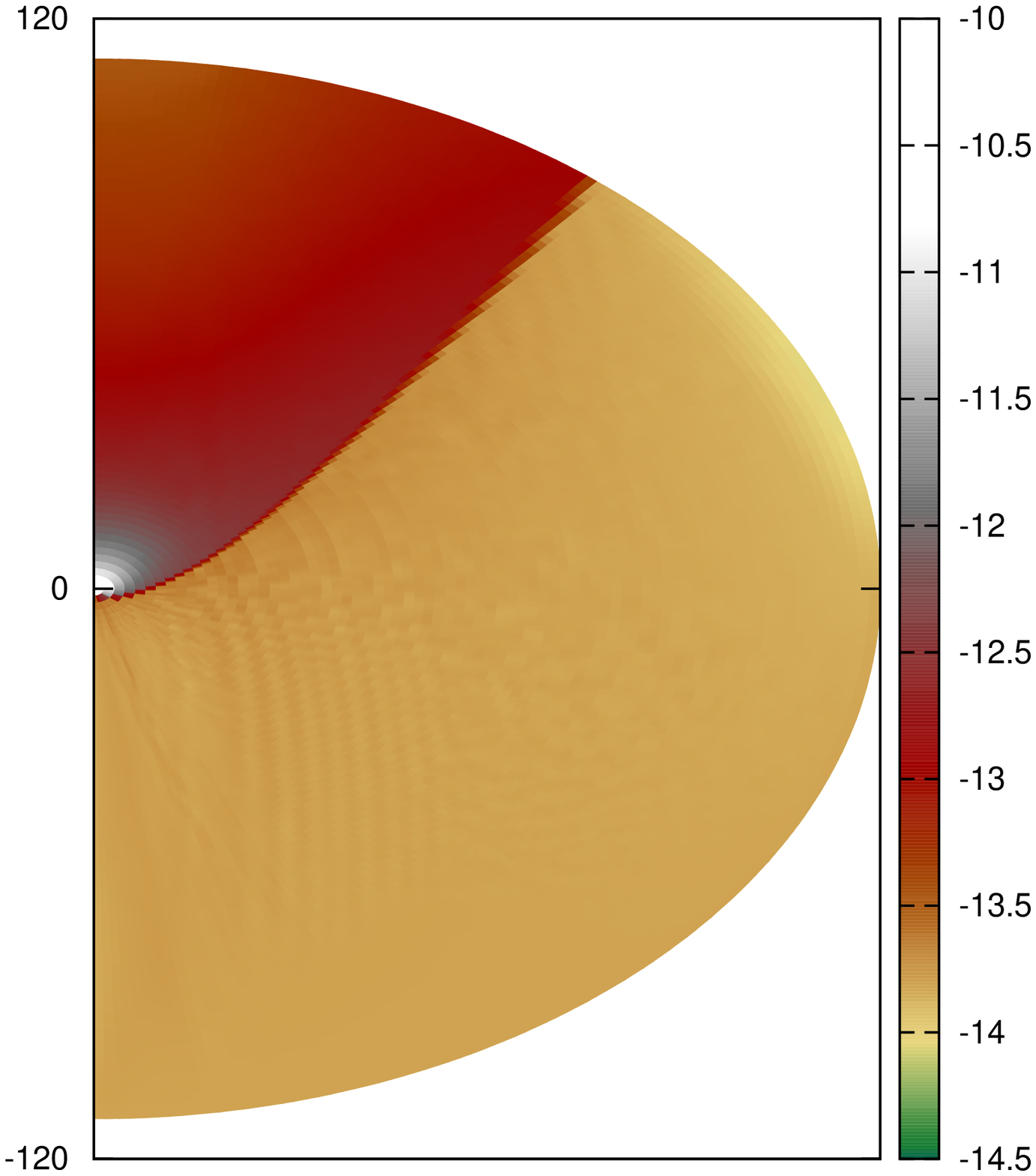}
\includegraphics[width=4cm]{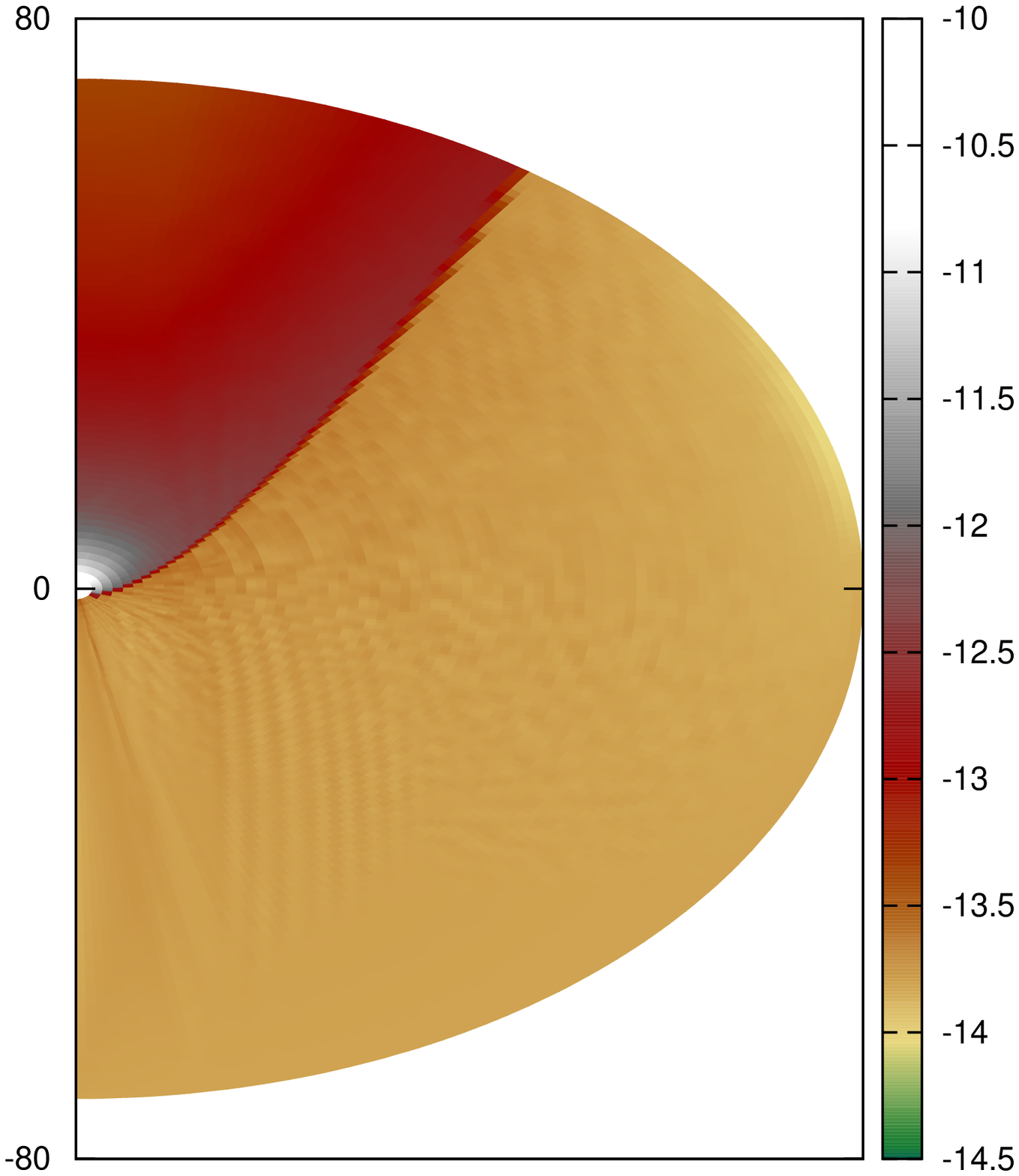}
\includegraphics[width=4cm]{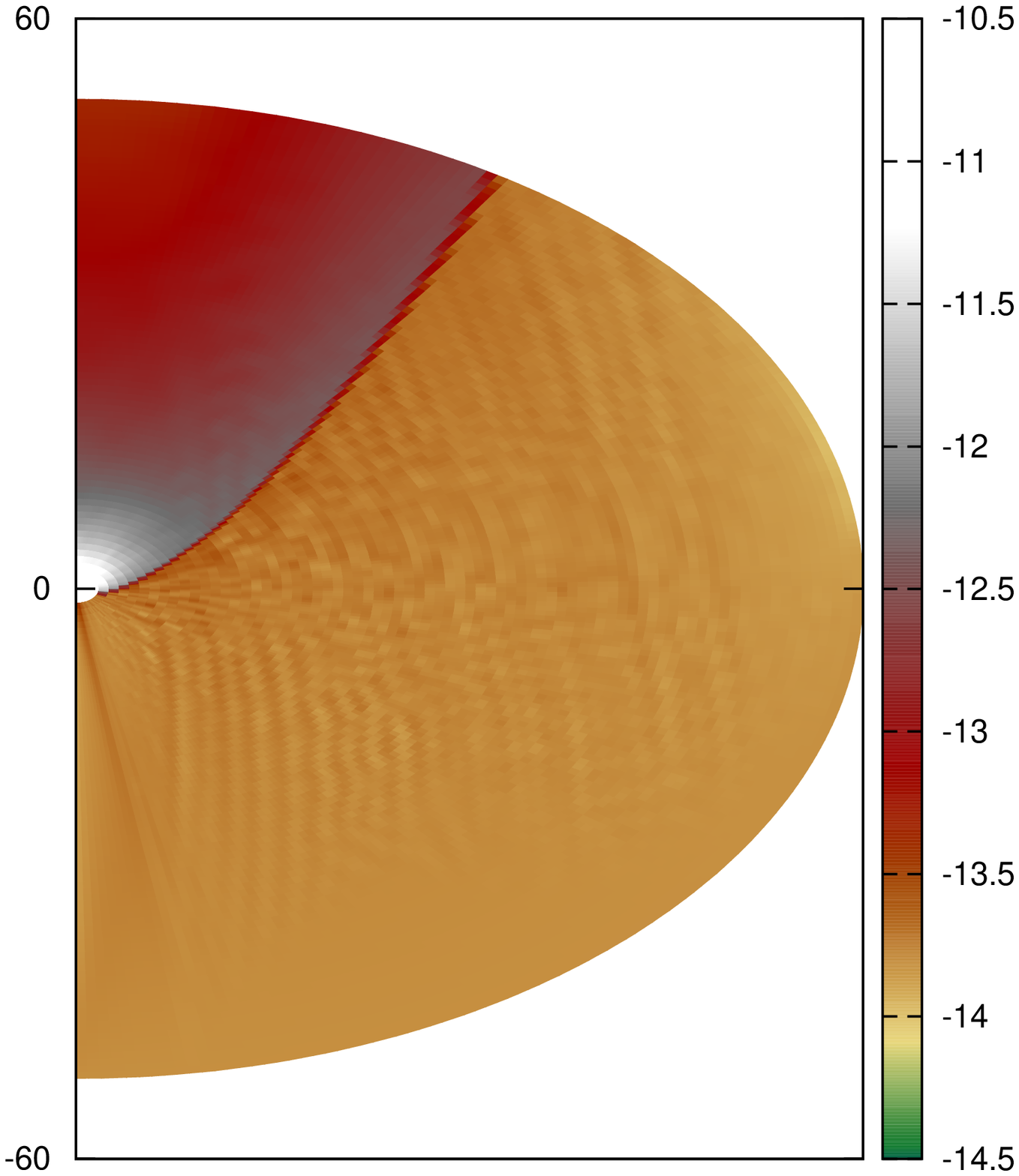}
\caption{\label{fig:Morpho} Logarithm of the density that illustrates the morphology of the shock cones in various scenarios when the system has achieved a nearly stationary regime. In the top panel we present the cases for the wind in models (from left to right) $M_{4a}$, $M_{3a}$ $M_{2a}$ and $M_{1a}$, whereas in the bottom for models $M_{4b}$, $M_{3b}$, $M_{2b}$ and $M_{1b}$ described in Table \ref{tab:InitialData}.}
\end{figure*}

In Fig. \ref{fig:Morpho}, we show the morphology for the different models we have considered in Table {\ref{tab:InitialData}}. As the parameter space is very large, we only chose two values for the adiabatic index $\Gamma=4/3,5/3$. Moreover, we only consider four initial wind supersonic velocities of the gas  at infinity corresponding to $2,3,4,5$ Mach. We then show that the use of penetrating coordinates does not change the well known morphology found in previous relativistic studies when using time-like internal boundaries, that is, the bigger the Mach number the smaller the shock cone angle, and also when $\Gamma$ increases the shock cone angle increases.

\begin{figure}
\includegraphics[width=8.0cm]{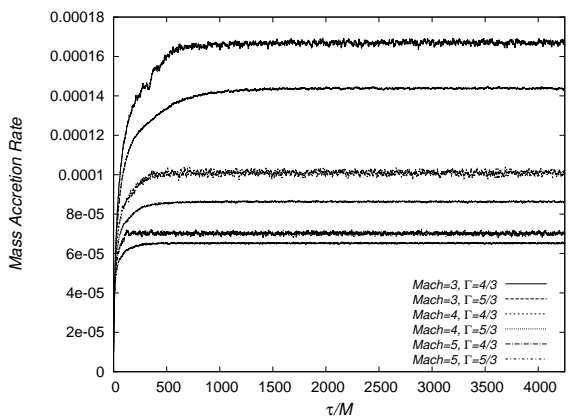}
\caption{\label{fig:MAccR} We show the mass accretion rate $\dot{M}_{acc}$ with respect to the proper time for different values of the Mach number and $\Gamma$. As we expect the accretion stabilizes faster when the Mach number and $\Gamma$ increase. This quantity is calculated in a detector located at $r=2.1M$, close to the event horizon. Despite the fact that we are calculating the mass accretion rate near the event horizon, no high oscillations show up and our measurements are accurate.}
\end{figure}

In Fig. \ref{fig:MAccR} the mass accretion rate for different values of the Mach number and $\Gamma$ is presented. When the Mach number and the adiabatic index increase, the system reaches the stationary regime faster. This has been already discussed in great detail by \cite{Fontaxi} and to us represents a test. On the other hand, some morphological effects of using horizon penetrating coordinates for non axial accretion and flip-flop behavior of the shock have been recently presented \cite{CruzLora2012}.

\begin{figure}
\includegraphics[width=8.0cm]{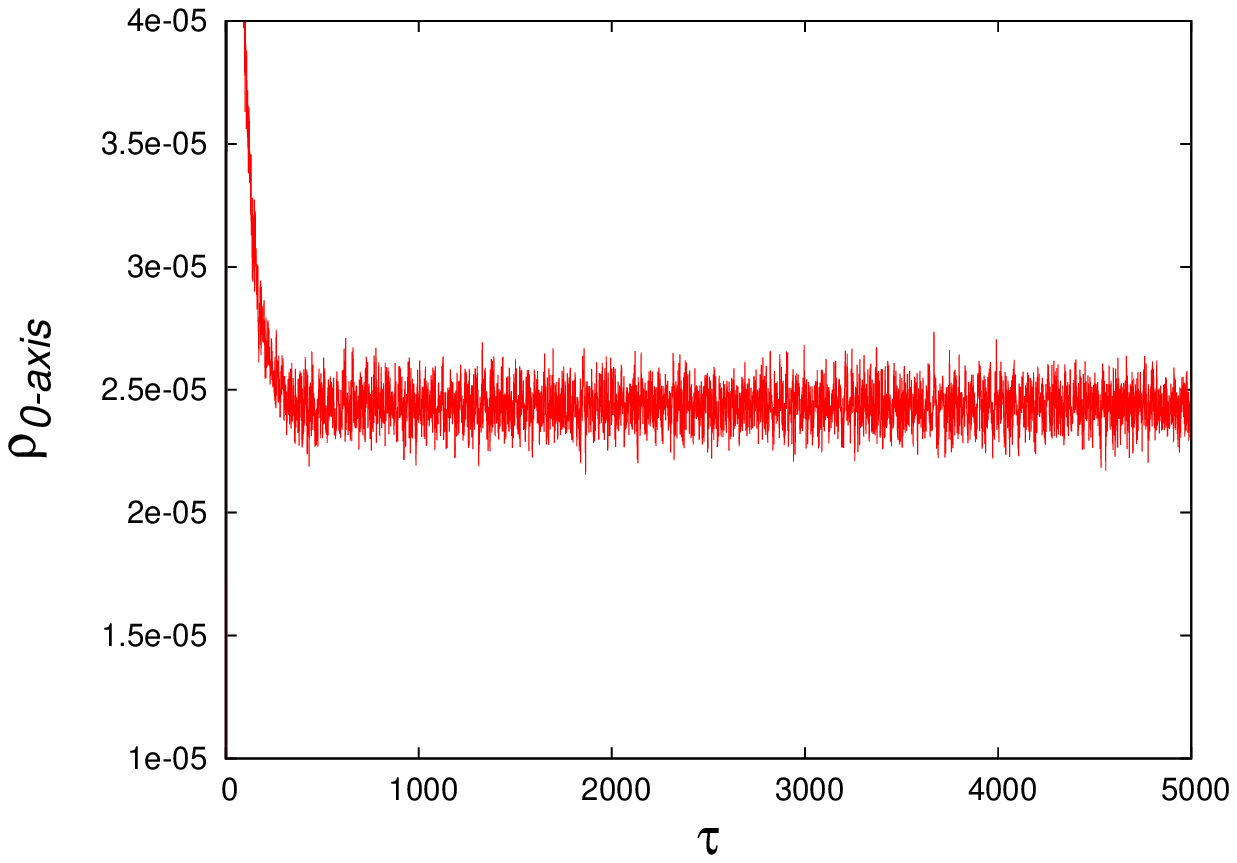}
\includegraphics[width=8.0cm]{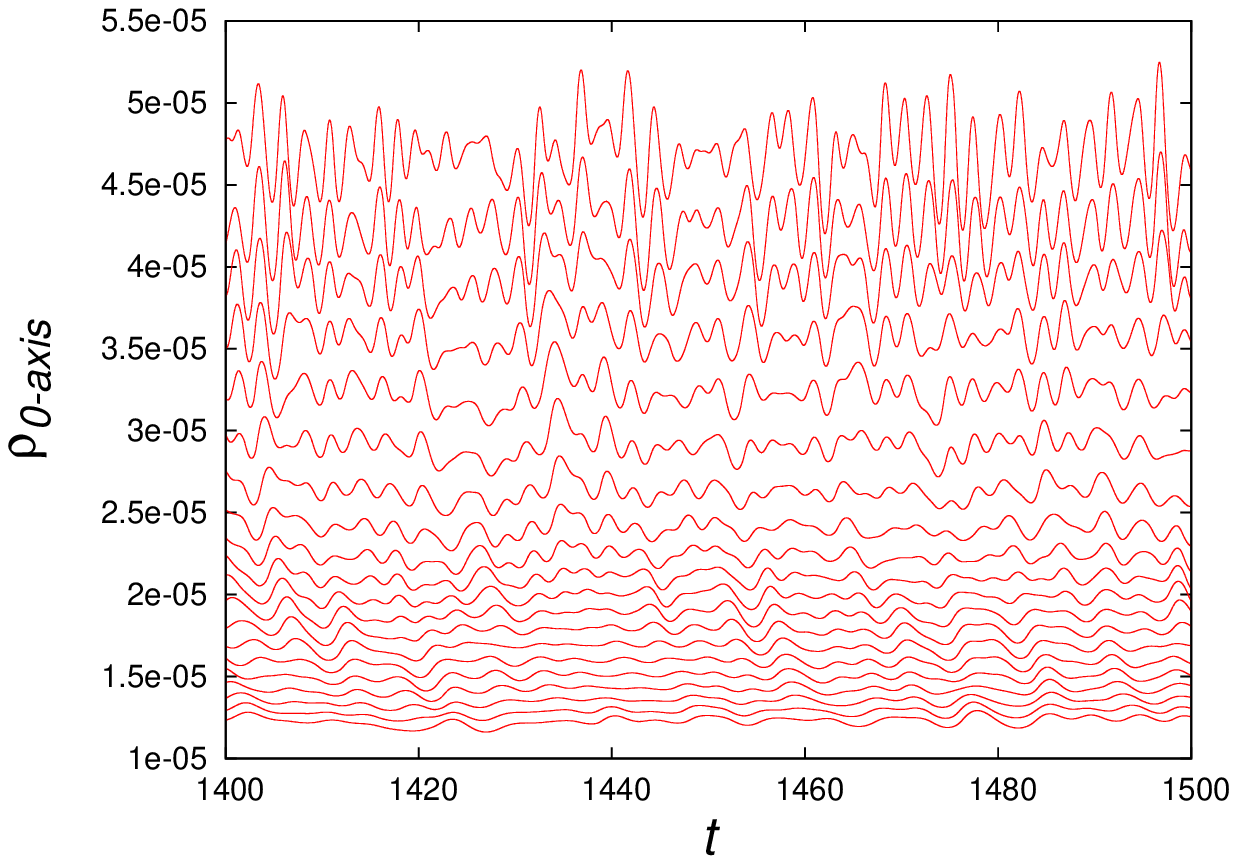}
\caption{\label{fig:rho_time} At the top we show how the density oscillates in time as measured by the detector at $12.1M$. In the second plot we show a zoomed time window showing the signals as measured by various detectors equally spaced being the first one located at $2.1$ and the farthest one at $16.1$; it can be seen the correlation among the signals measured by successive detectors; it is also shown that high frequency modes that are measured by detectors close to the hole are not seen by farther detectors. This result is complemented with power spectra below.}
\end{figure}

Once we have described some of the properties of the shock cone we measure the density at various  points as a function of time. What happens is that the density vibrates as shown in Fig. \ref{fig:rho_time}. In our approach we define the inner boundary inside the black hole where the light cones are oriented toward the black hole singularity and are open as opposed to the BL coordinates that shows light cones that are closed when approaching the event horizon. Thus, the fluid particles move inside the black hole naturally. 

The fact that the shock cone vibrations are not associated to numerical artifacts, allows us to explore, as a toy astrophysical model, a possible relation between such oscillations inside the shock cone and QPOs frequencies.

\begin{figure}
\includegraphics[width=8.0cm]{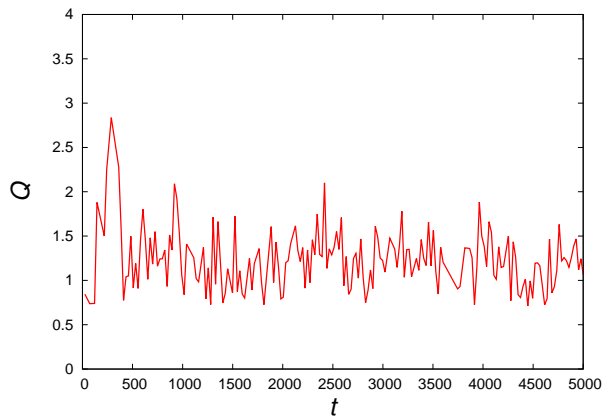}
\caption{\label{fig:conver} Self-convergence of the $L_2$ norm of the density along the positive $z$-axis, where the shock cone forms. The ratio among successive resolutions is 1.5, and the convergence factor is defined by $1.5^Q$. The accuracy of our algorithms is expected to be first order due to the development of shocks.}
\end{figure}

\section{A possible model of QPO sources}
\label{sec:QPOs_results}

A rather appealing potential application of the vibrations of the density of the gas within the shock cone is related to the QPO type of emission. In \cite{Donmez} it is found the first work studying the possibility that the vibrations of the density within the shock cone could be associated to QPO frequencies. Some differences in our approach are that in our analysis uses penetrating coordinates in the description of the black hole, whereas in \cite{Donmez} they are using singular coordinates and our study is  axisymmetric whereas in \cite{Donmez} slab symmetry is used. With these tools we ask whether the most powerful peaks are approximately within the range of frequencies observed for QPOs in terms of the black hole mass. 

It is worth to mention that in our relativistic approach, aside of the restrictions of the realistic velocity of the wind, we do not include non-adiabatic cooling and heating processes since this is a first step toward an exhaustive and realistic analysis in the general relativistic regime truly allowing the gas entering the black hole. The first steps on that direction involving radiation terms have been recently published \cite{Zanotti,referee2}.

We now analyze the oscillations as measured at various distances from the black hole. In Fig. \ref{fig:rho_time}  we show various aspects of the oscillations. Firstly we show the oscillations as measured by a detector located at distance $12.1M$ in proper time; it can be seen that after an initial transient the rest mass density approaches a nearly stationary regime; secondly we show a time window showing the signal as measured by detectors located at various distances, ranging from 2.1$M$ to $16.1M$; from this plot we learn that detectors closer to the black hole measure oscillations dominated by high frequency modes, whereas detectors far from the black hole measure signals dominated by lower frequency oscillations; moreover, this plot shows that signals among detectors are well correlated and that we are not measuring only numerical noise. We support our calculations with the a self-convergence test of the density in Fig. \ref{fig:conver}.

\begin{figure}
\includegraphics[width=4.0cm]{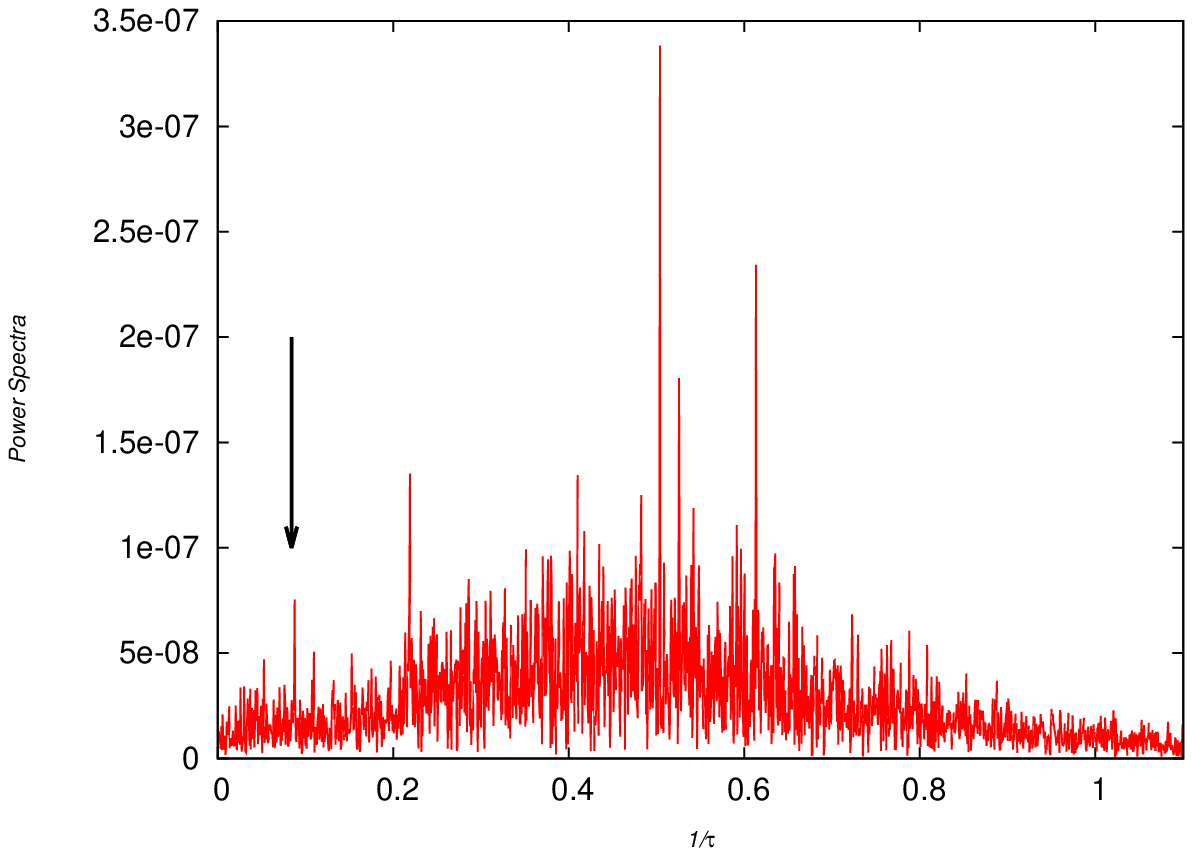}
\includegraphics[width=4.0cm]{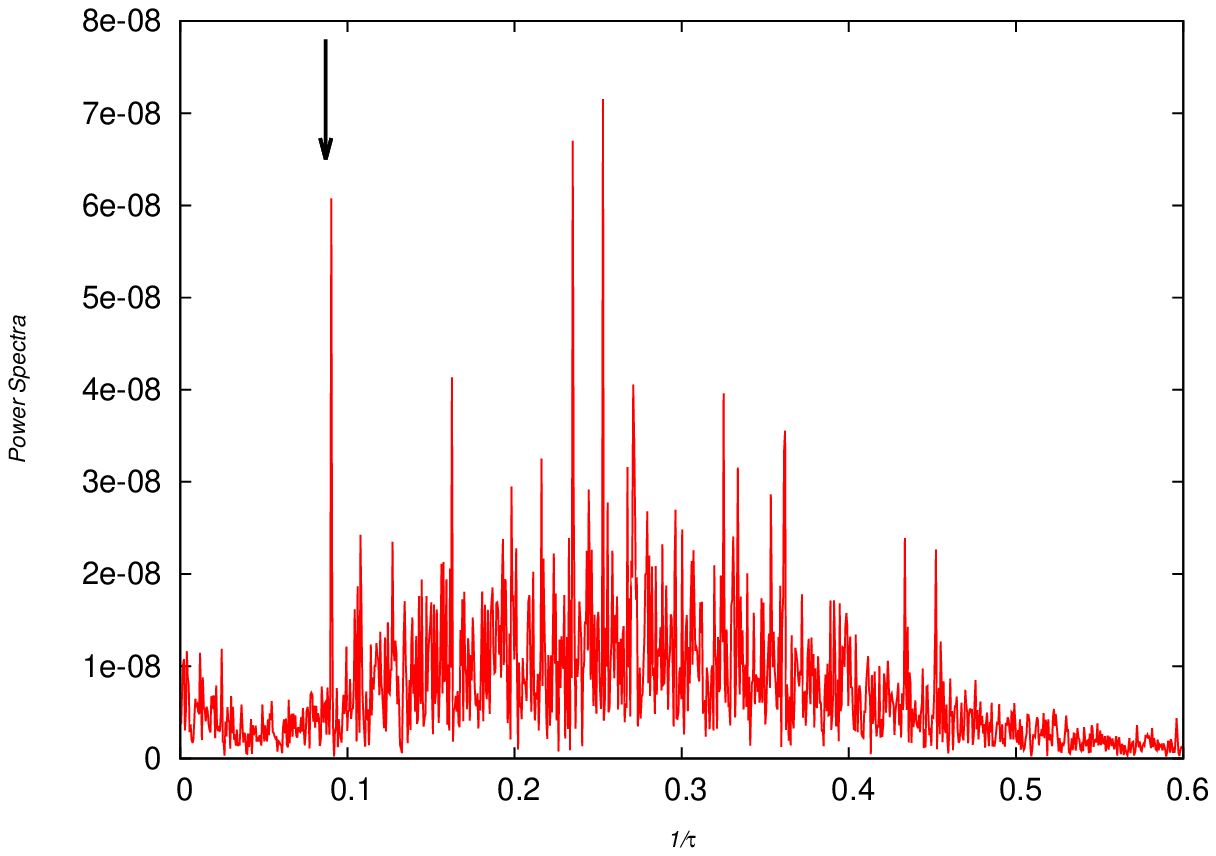}
\includegraphics[width=4.0cm]{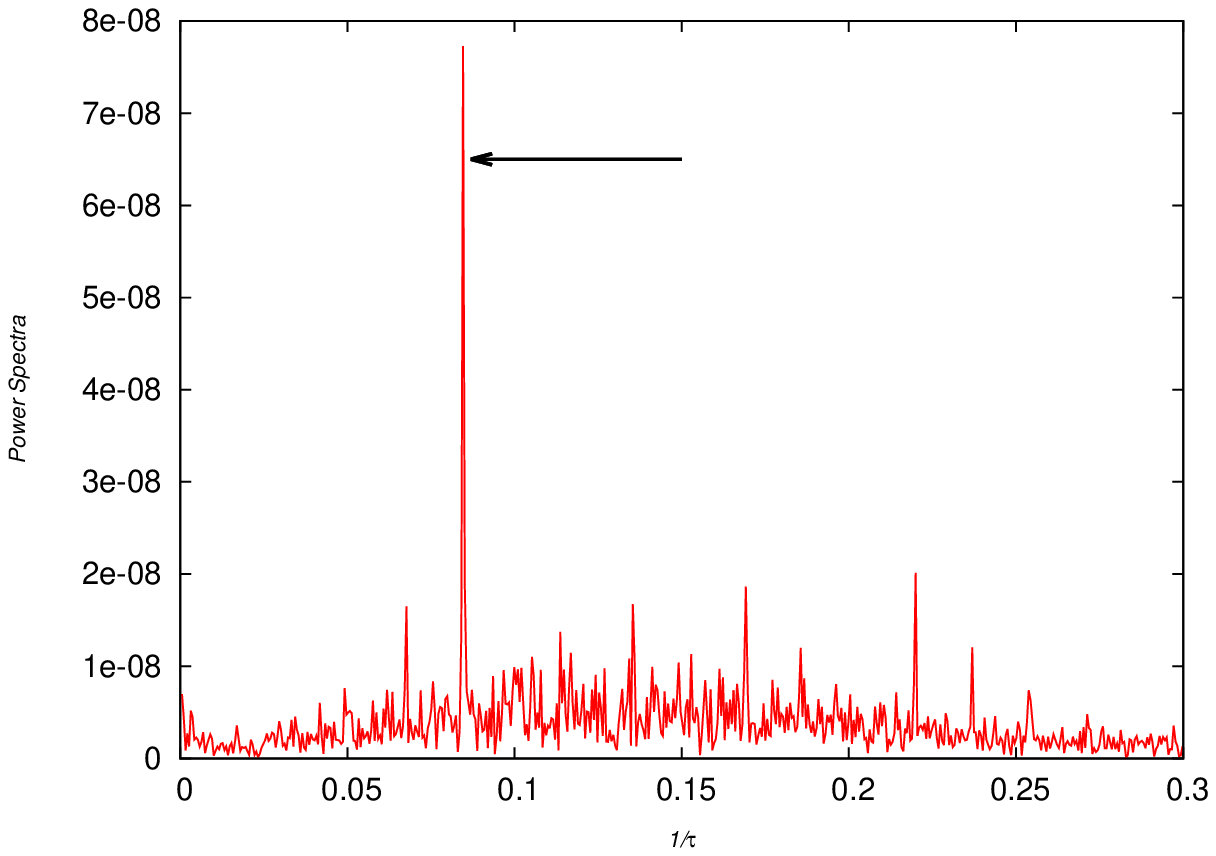}
\includegraphics[width=4.0cm]{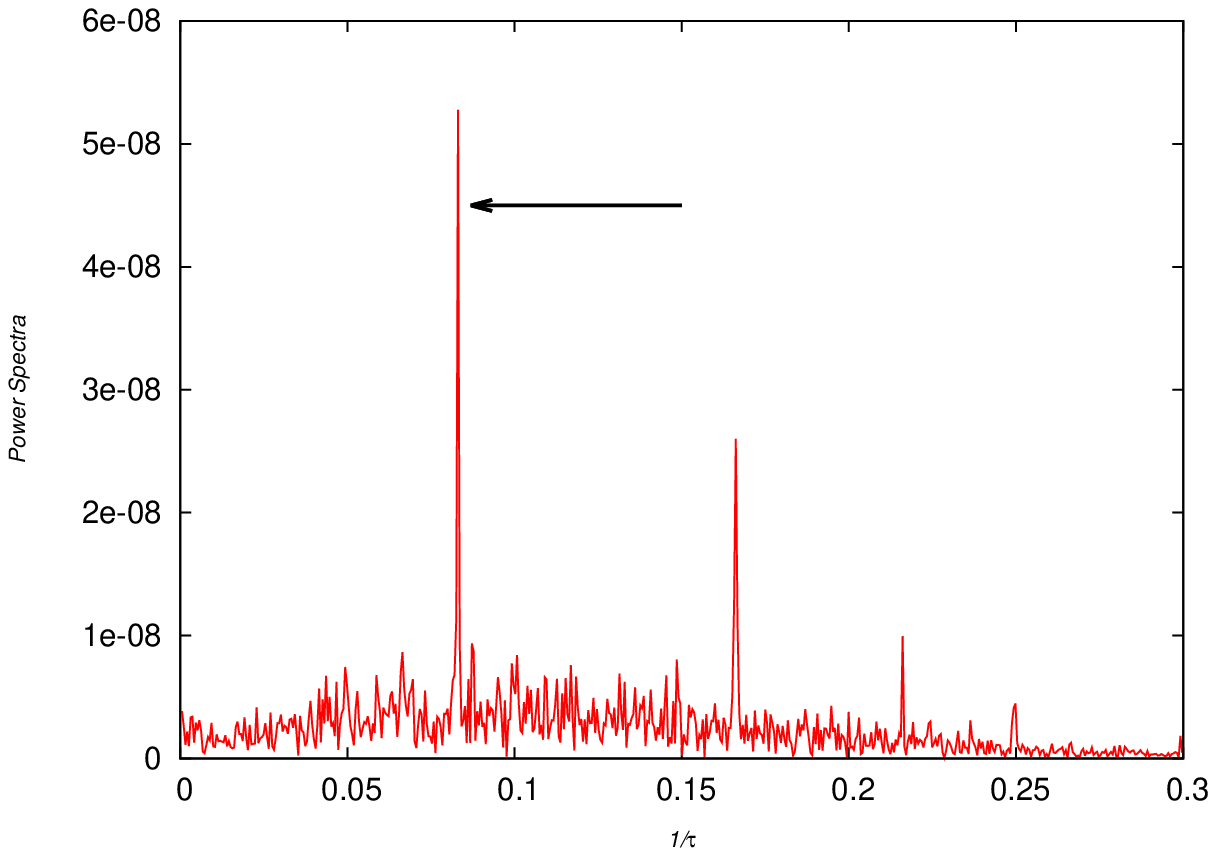}
\caption{\label{fig:QPOS} Spectrum of density oscillations in proper time for the model $M_{1b}$ as measured at different detectors located at: (Top-left) $r=2.1M$, (top-right) $r=6.1M$, (bottom-left) $r=12.1M$ and (bottom-right) $r=16.1M$.}
\end{figure}

In order to interpret the oscillations of the density we perform a Fourier transform of the signals measured by different detectors like those in Fig. \ref{fig:rho_time} using the proper time at the location of each detector. The results appear in Fig. \ref{fig:QPOS}, where the high frequency modes are shown to dominate near the black hole and do not appear far from the black hole. We concentrate in modes that may be global and detected in a big part of the domain as suggested by the analysis in \cite{Donmez}. As a particular case, we point out with an arrow in Fig.  \ref{fig:QPOS} a frequency peak that appears in all the detectors. That it appears in all our detectors is an indication that the corresponding mode is global and this is the reason why we choose it to correspond to the frequency in our analysis. In all the cases in our parameter space the spectra show the presence of such mode. This is the reason why in all the cases studied we consider the spectrum as measured with a detector located at $r=12.1M$ where the spectrum is very clean and the most powerful peak corresponds precisely to this mode.

We then proceed to explore the chances that such frequency lies within observed QPO frequency ranges. Our parameter space covers the models described in the Table for various values of the speed of sound and velocity of the wind. For each case we locate the most powerful peak as measured by the detector at $r=12.1M$ and obtain a proper frequency of oscillation of the gas within the shock cone. In this paper we push the wind velocity values down to $0.05c$. We consider that the smallest velocity a wind requires to form a shock cone is that of the speed of sound. However, when the speed of sound is small, the accretion radius is big in terms of the length scale of the black hole (see the bold face parameters in the Table and eq. (\ref{eq:racc})) and this imposes a computational limitation on the set of parameters that can be used to complete a simulation, because one requires not only a big spatial domain, but also enough resolution near the black hole and in the whole domain.

\begin{figure}
\includegraphics[width=8.0cm]{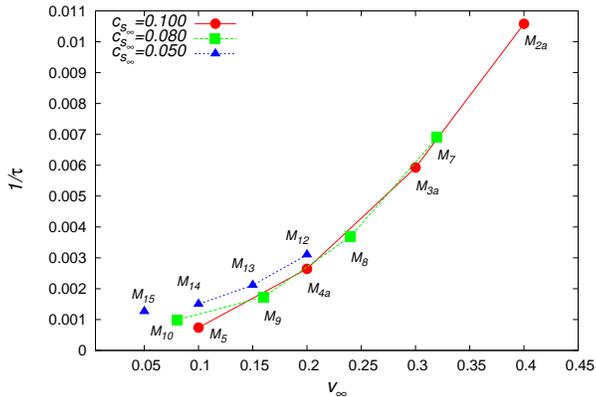}
\caption{\label{fig:extra2mach1} We show the frequencies of oscillation of the rest mass density in the shock cone in terms of the wind velocity for different values of the speed of sound.}
\end{figure}

In Fig. \ref{fig:extra2mach1}  we show the frequencies obtained for the simulations in our Table. Circles correspond to the results obtained from simulations considering a sound speed $c_{s \infty}=0.1$.  Analogously, squares and triangles indicate our results with $c_{s \infty}=0.08$ and $c_{s \infty}=0.05$ respectively, being the model $M_{15}$ an extrapolation to Mach 1 given the numerical domain restrictions described.

\begin{figure}
\includegraphics[width=8.0cm]{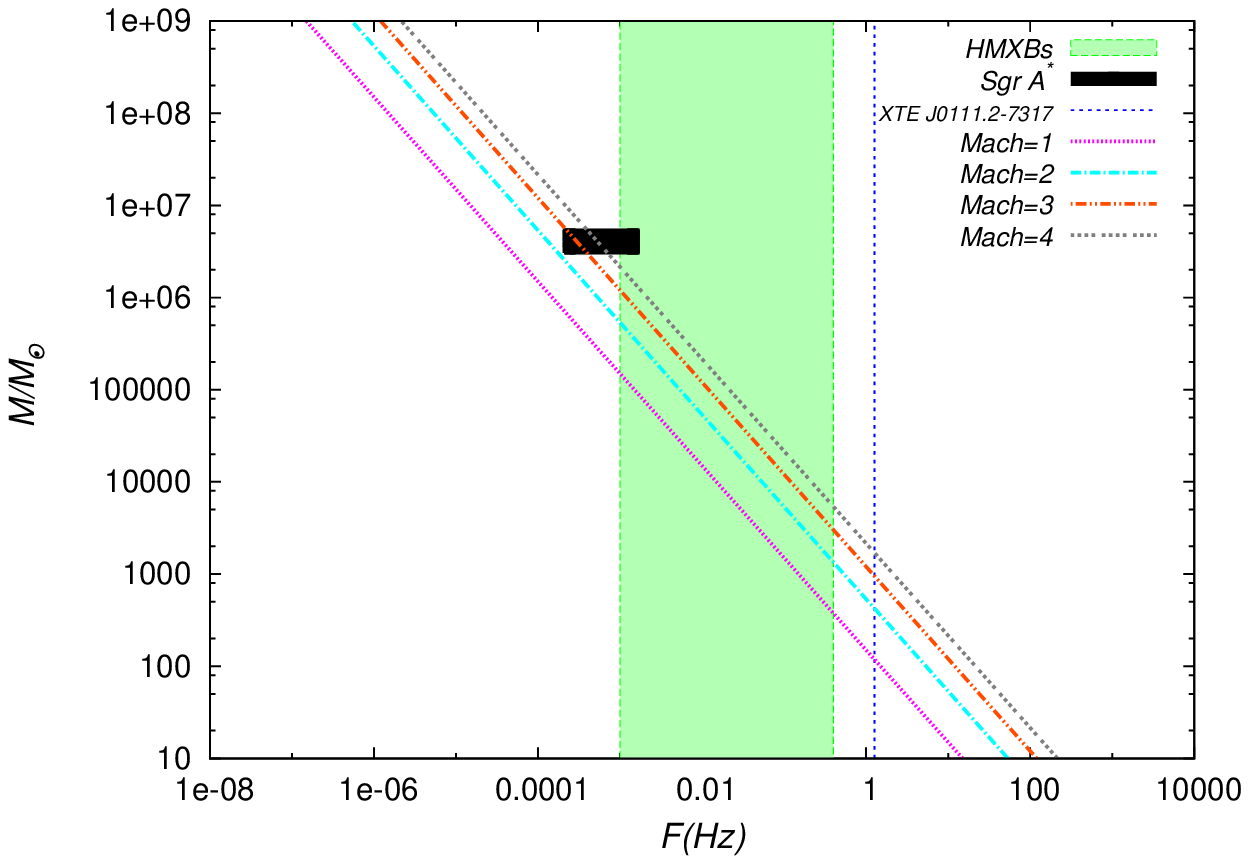}
\includegraphics[width=8.0cm]{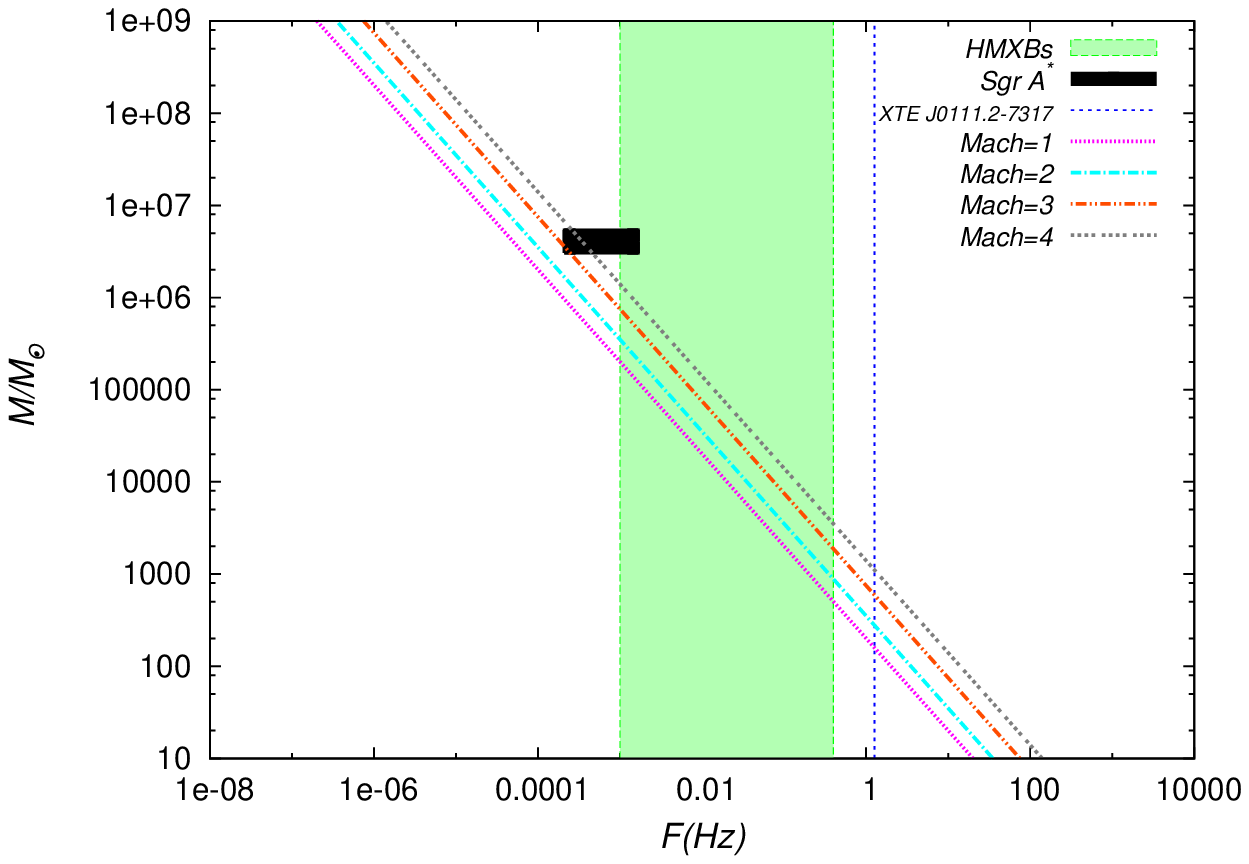}
\includegraphics[width=8.0cm]{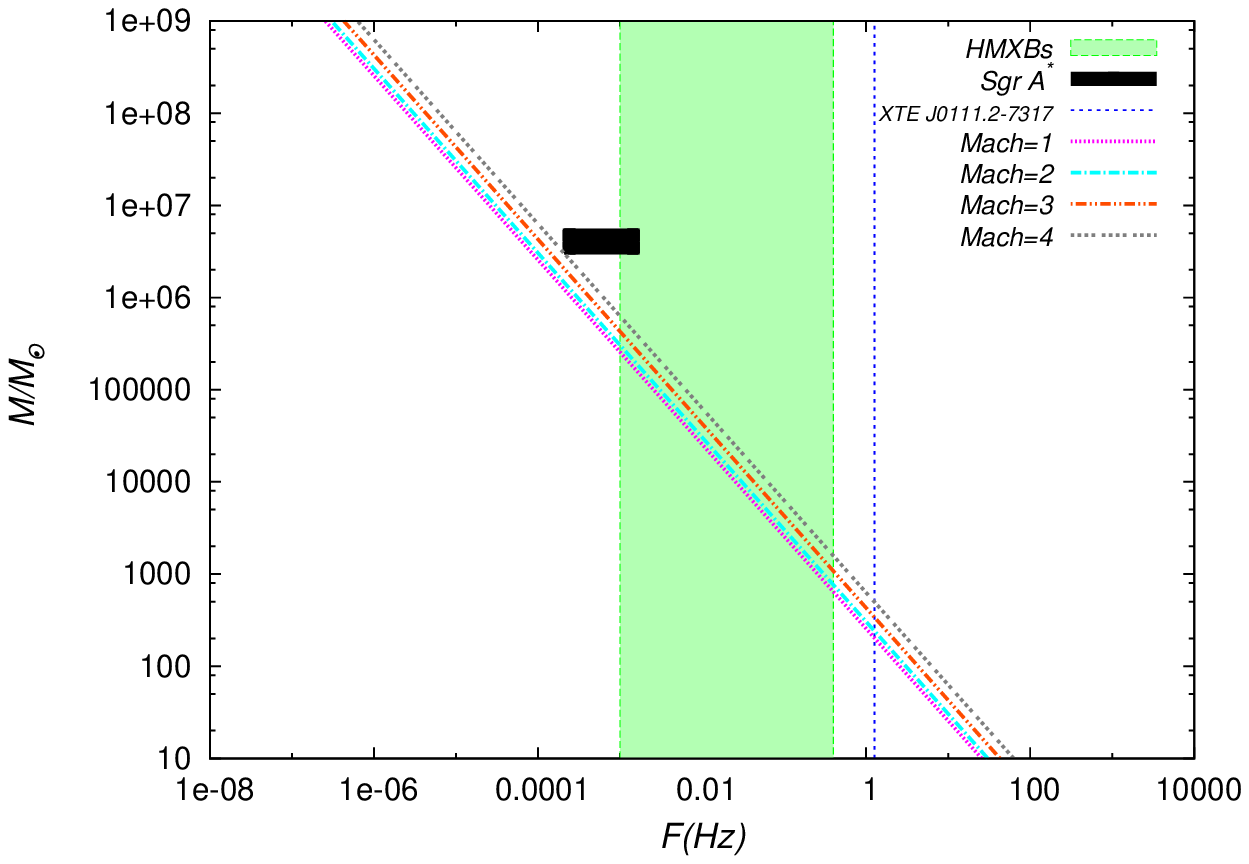}
\caption{\label{fig:MvsF} We summarize our study of the parameter space. From top to bottom, the speed of sound used is $c_{s_{\infty}}=0.1,~0.08,~0.05$. We also show a vertical line indicating a special frequency of 1.27Hz associated to HMXBs. We also indicate a shadowed region corresponding to HMXBs in the range of 1mHz - 400mHz. The black box corresponds to the black hole mass and frequency range observed in Sgr A${}^{*}$. We show the case for $\Gamma=4/3$.}
\end{figure}

In Fig. \ref{fig:MvsF} we compare the frequency window we find in all our simulations with observational results. We ask about the value of our parameters needed for our frequencies to lie within the range of those observed from QPO sources. Each point in each of the diagonal lines defines a proper frequency of the peak we have selected in physical units for several different scenarios, in terms of the black hole mass. Each one corresponds to a different value of the velocity of the wind in units of the speed of sound. We obtain these lines by simply scaling our results, which were calculated in units of $M=1$.
In the first plot we show the frequencies in terms of the mass of the black hole for various values of $c_{s \infty}$. For each value of the speed of sound (each plot in Fig. \ref{fig:MvsF}) we see that the bigger the velocity of the wind the bigger the frequency allowed. On the other hand, the three different panels show that the bigger the value of the speed of sound $c_{s \infty}$ the bigger the frequency as well.

In order to compare with observations, we include in Fig. \ref{fig:MvsF} the masses and frequencies corresponding to the case of Sgr A${}^{*}$, where the mass of the supermassive black hole is $M=4.1\pm0.6 \times 10^{6}M_{\odot}$ \cite{Ghez} and the frequency range was the one reported in \cite{Aschenbach,Abramowicz}. We also show a shadowed region corresponding to the range of masses and frequencies associated to high mass X-ray binaries (HMXBs). 

Based on our numerical results in Fig. \ref{fig:MvsF} we notice that for the range of velocities  analyzed, for the model with $\Gamma=4/3$ and various wind velocities, the frequencies of the density oscillations and mass of the black hole for Sgr A${}^{*}$, lie partially within that of observations. The trend observed between panels in Fig. \ref{fig:MvsF} indicates that when $c_{s \infty}$ increases, a bigger range of our parameters contain Sgr A${}^{*}$.

Fig.  \ref{fig:MvsF} also shows a shadowed box corresponding to QPOs between $1mHz$ to $400mHz$, observed in the spectra of HMXBs. Some of these measurements are associated with regions where it is believed a black hole exists, $e.g$ Cyg X-1 \cite{Liu}. As we can see, this figure  predicts a variety of possible configurations for the observed frequencies. For example,  the allowed mass of the black hole that is hosted in the observed regions can run from hundreds to millions of solar masses for all the models presented here.

The vertical dashed line in Fig. \ref{fig:MvsF} corresponds to another QPO source of $1.27Hz$ \cite{Kaur}. Consistent black hole masses would be of about $10^2$ to $10^3 M_{\odot}$.

An exhaustive further study of the parameter space, would include information on the equation of state, additional conditions involving non-adiabtic and radiative processes like in \cite{Zanotti,referee2} in more realistic models. However it would be interesting as a first step the use of realistic wind velocities in simulations, which may involve the use of fish-eyed type of coordinates to describe the background space-time or the use of compactified coordinates with special slicing conditions on the space-time that penetrate the event horizon and approach infinity at once, that would allow us to study a bigger physical domain in terms of the accretion radius, using modest computational resources, and associating bigger resolution in the regions near the black hole, as done in other already studied cases of the propagation of scalar waves onto black hole space-times where the future null infinity can be contained in the numerical domain \cite{CMC1}.

\section{Conclusions and discussion}
\label{sec:conclusions}

We have studied numerically the  axisymmetric relativistic Bondi-Hoyle accretion of a supersonic ideal gas onto a fixed Schwarzschild background spaceÐtime described using horizon penetrating coordinates. Our results reproduce the morphology of the distribution of the rest mass density of the gas obtained in the Newtonian regime and in previous relativistic studies and verify that a stable shock cone forms.

With our approach we measured the consistency in the rest mass density within the shock cone, which is an indication that the excision implemented inside the black hole works fine with accuracy and convergence. Our treatment contributes to the solution of the Bondi-Hoyle accretion problem at the scale of the accretor scale length. This will provide a better understanding of the gas dynamics near the black hole. The restriction of this local type of study in astrophysical scenarios at the moment is that one needs the use of a small accretion radius in order to cover a sufficiently large numerical domain, which in turn implies that the velocities that can be studied are higher than those observed or estimated by theoretical models.

One of the properties of the accretion of a relativistic supersonic wind is that the density within the shock cone vibrates. We explore the possibility that such vibrations show frequencies within the range of those observed in QPO sources. Our approach is the first one for axisymmetric flows, in the relativistic regime, on a space-time corresponding to a black hole, which additionally is described in coordinates that truly allow the gas flow inside the black hole horizon.

\section{Appendix: tests of the code}

In order to illustrate how our implementation handles the evolution of a gas in a relativistic scenario, in this appendix we present standard tests. In all of them we use the minmod to reconstruct the variables and the HLLE flux formula, as we do in the simulations presented in this paper.

\begin{figure*}
\includegraphics[width=5.5cm]{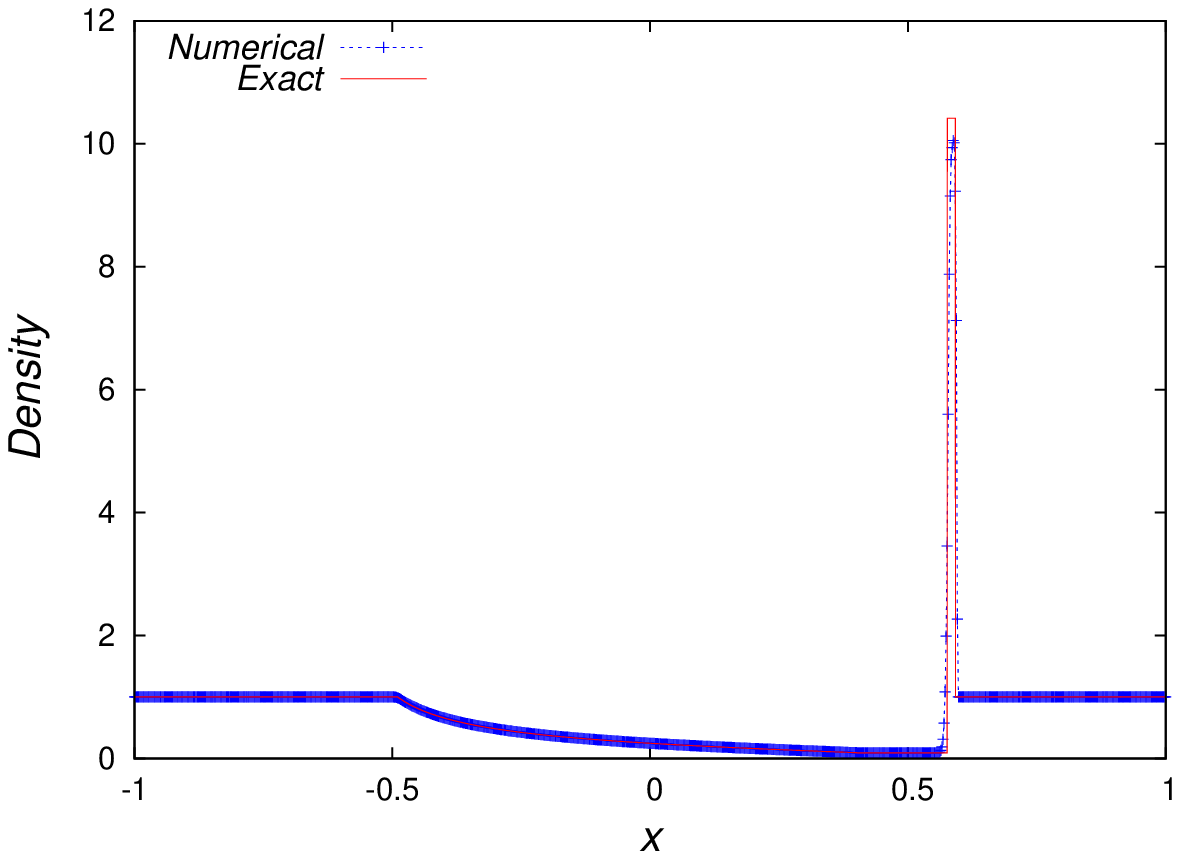}
\includegraphics[width=5.5cm]{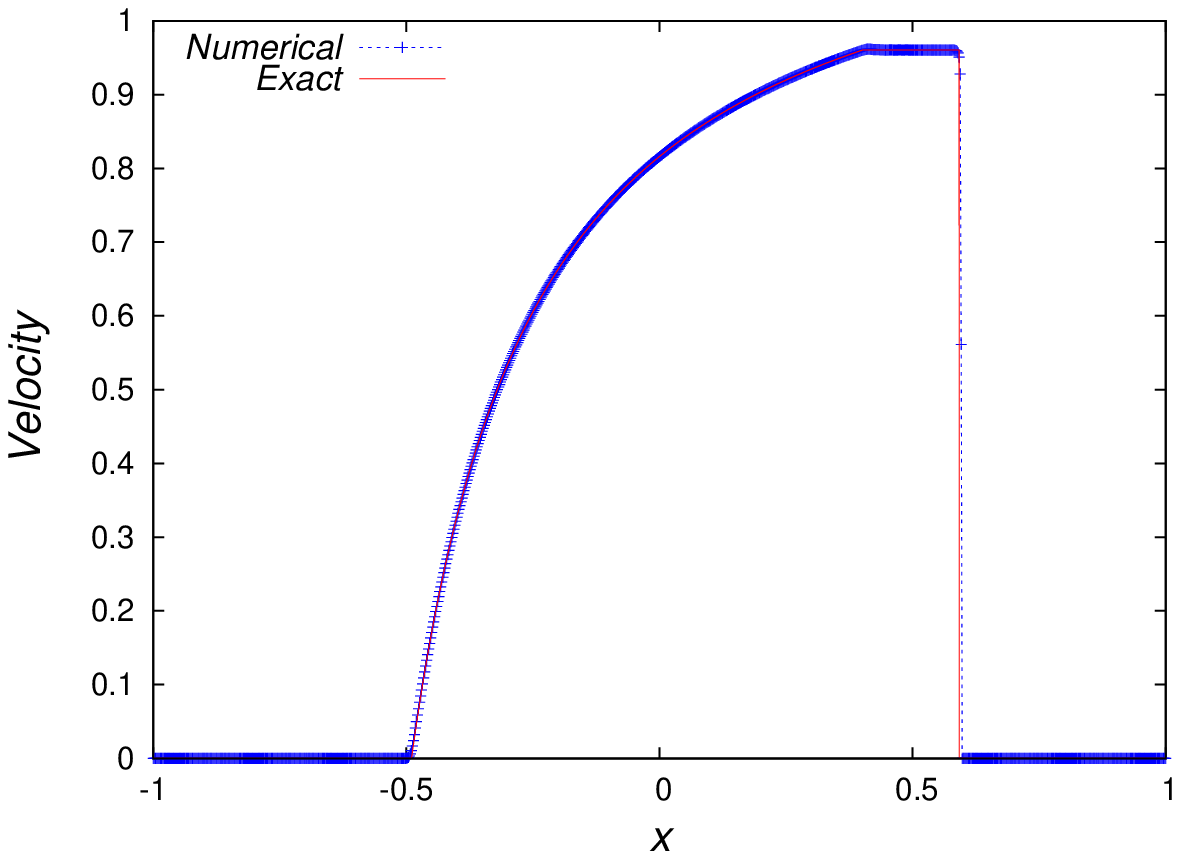}
\includegraphics[width=5.5cm]{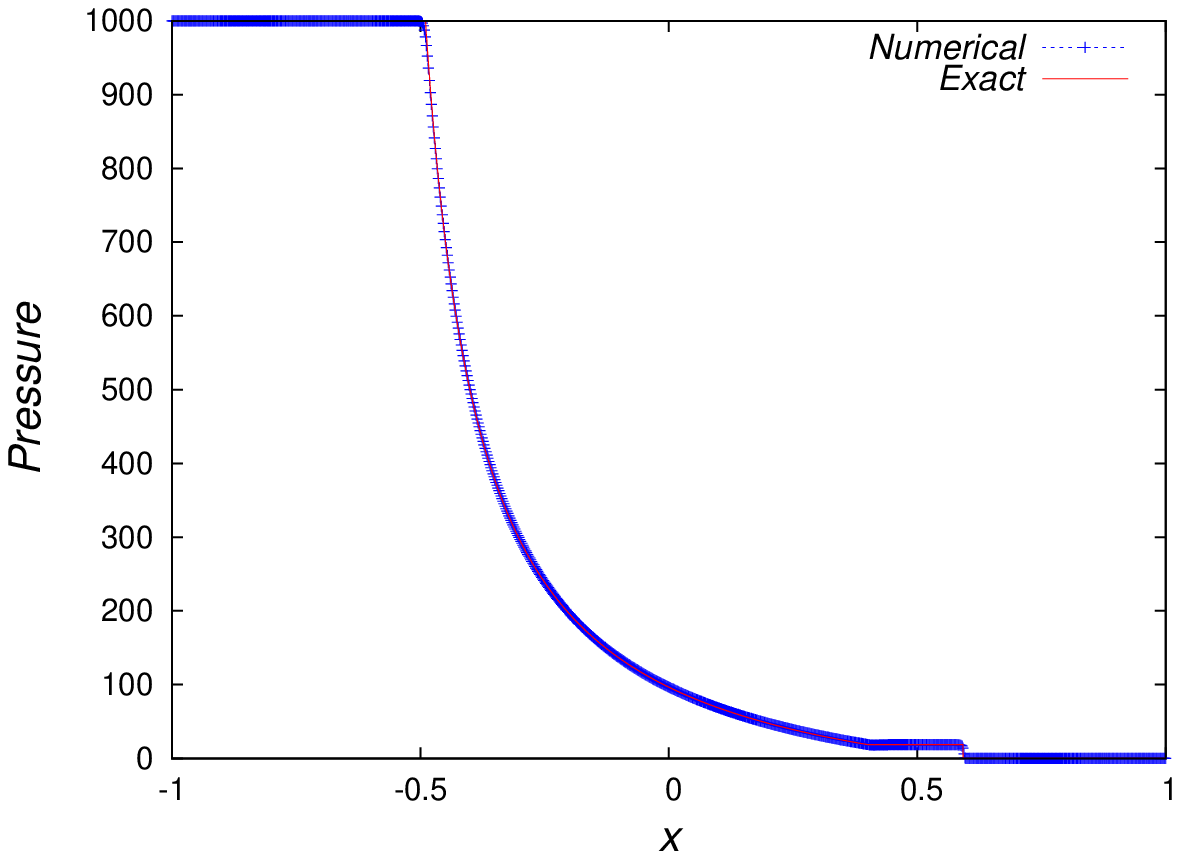}
\caption{\label{fig:shock-tube} We show the gas variables at time $t=0.6$ and the exact solution. As a global convergence test involves regions where constant states are involved, the velocity of the shock, and the shell of density at the front shock are important quantities to monitor. We resolve the shock with only a few points and the velocity at the time of the snapshot is very accurate. We use for this test 1000 cells and $\Gamma=5/3$.}
\end{figure*}

{\it Test 1, 1d tests.} The basic test a hydrodynamical code has to satisfy Sod's shock tube problem. We choose the initial parameters of the commonly called blast wave problem, with standard parameters as in \cite{MartiLRR}. The most stringent test has to do with the density profile at the front shock and the velocity of the shock. In Fig. \ref{fig:shock-tube} we show our results and contrast them with the exact solution, which we calculate following \cite{MartiLR}. The initial conditions are $\rho_L = \rho_R =1.0$, $p_L=1000$, $p_R=0.01$, $v_L=v_R=0$.

\begin{figure*}
\includegraphics[width=5.5cm]{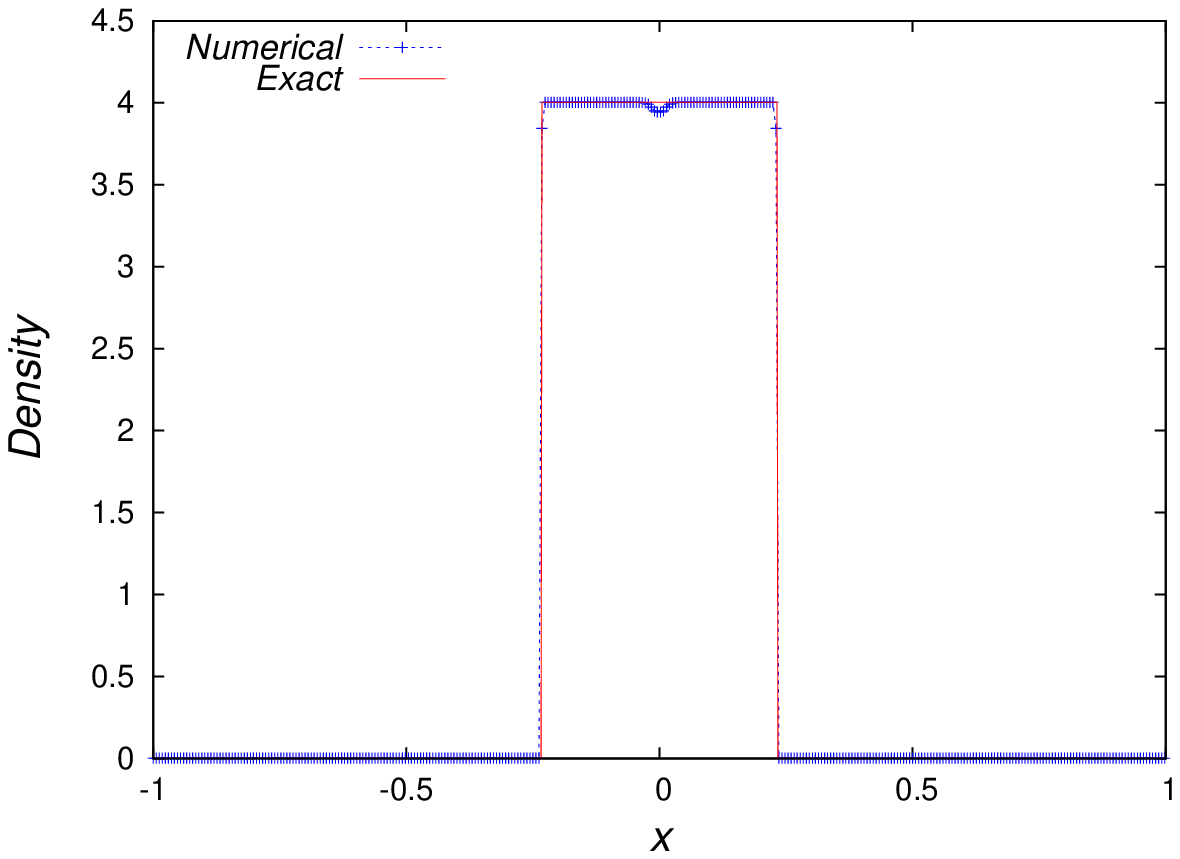}
\includegraphics[width=5.5cm]{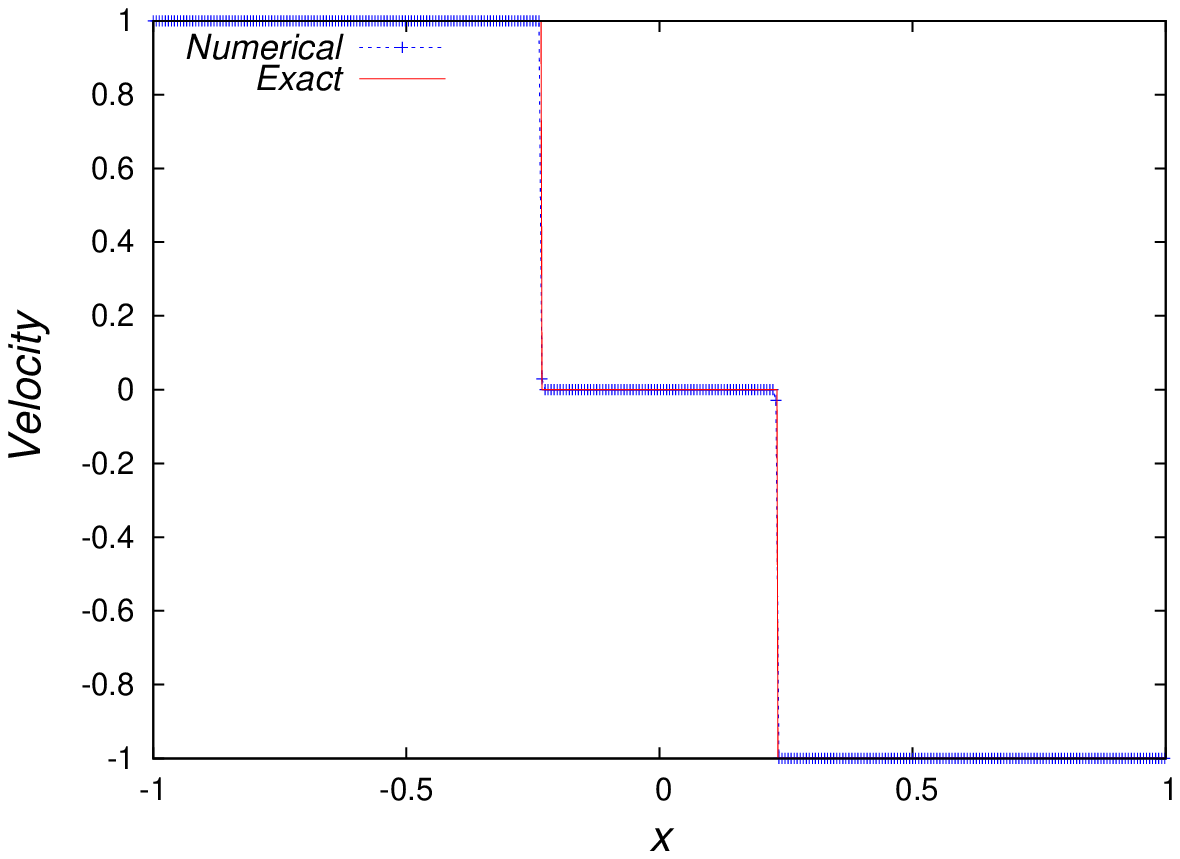}
\includegraphics[width=5.5cm]{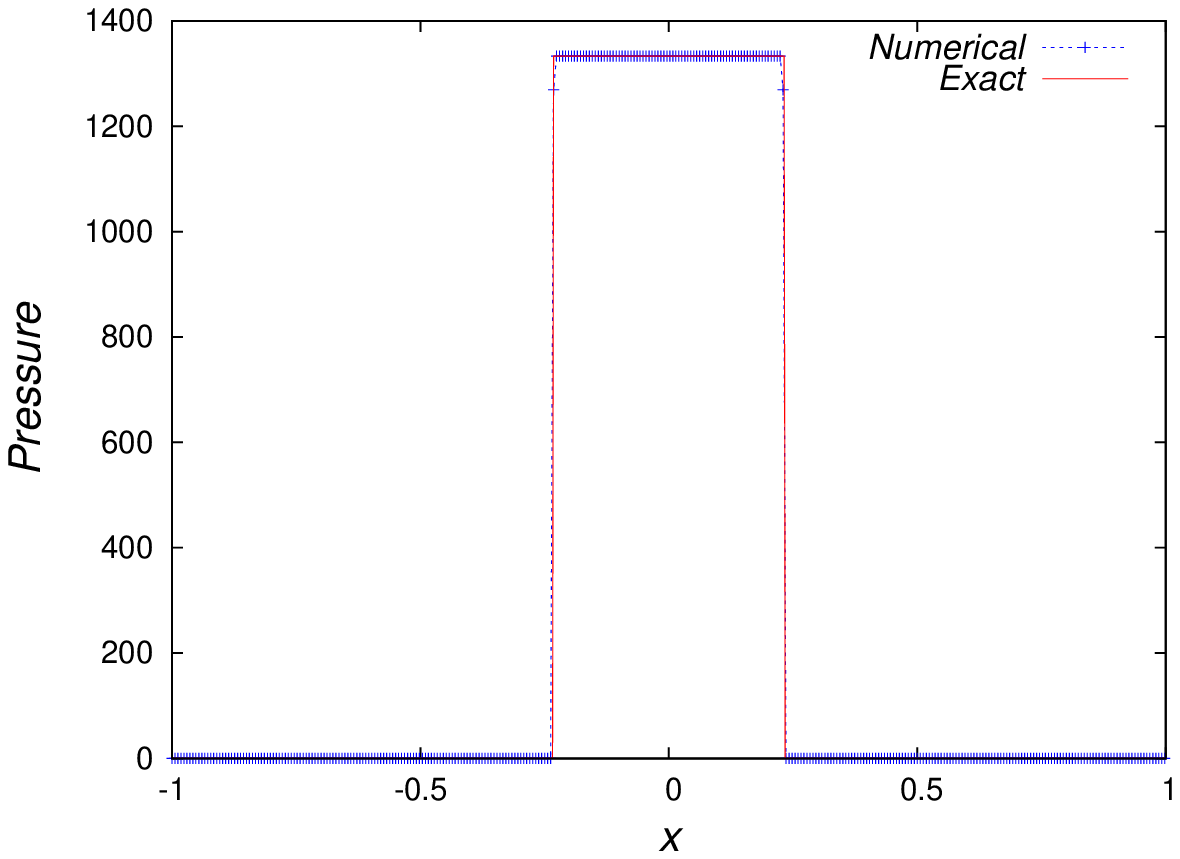}
\caption{\label{fig:shock-tube2} This test shows that our implementation is capable to track the evolution of velocities that correspond to Lorentz factors of the order of $\sim 1000$. The snapshot is taken at $t=0.7$. In this test we use $\Gamma=4/3$. Again we use 1000 cells.}
\end{figure*}

In order to test our implementation under a more extreme condition, we present what we call a head-on stream collision, which produces Lorentz factors of the order of 1000. The initial conditions on the initial state are: $\rho_L=\rho_R=0.0009999998749$, $p_l=p_R=3.33333258\times 10^{-9}$, $v_L=-v_R=0.9999995$, which are consistent with the wall-shock test in \cite{Riffert}. The numerical results compared with the exact solution are shown in Fig. \ref{fig:shock-tube2}.

\begin{figure}
\includegraphics[width=8.0cm]{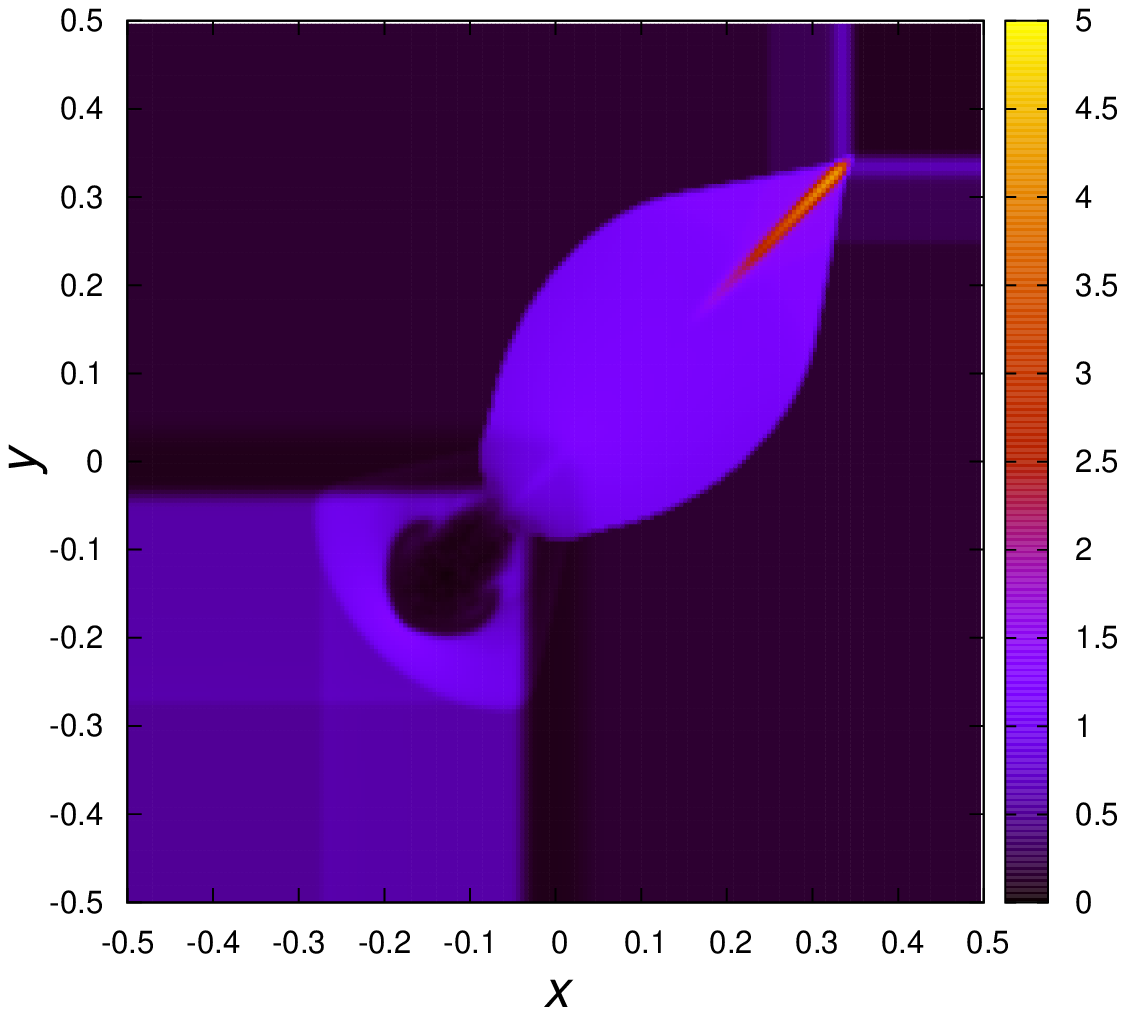}
\caption{\label{fig:2dshock-tube} Logarithm of the rest mass density  for the 2D relativistic Riemann problem at $t=0.35$. We illustrate the morphology developed by the process. We use 1200$\times$1200 cells to cover the domain.}
\end{figure}

{\it Test 2, 2d shocks.} Our code handles the topology of cartesian coordinates because we use excision, which allows us to show standard 2d tests. The initial conditions are constructed in such a way that a square domain is divided into four equal regions and where the variables start with the values shown in Table \ref{tab:2dshock}. A snapshot at $t=0.35$ is shown in Fig. \ref{fig:2dshock-tube}. We show the morphology as done in standard tests \cite{tests1,tests3}.

\begin{table}
\begin{tabular}{|c|c|}\hline
top-left & top-right\\
$\rho=0.1$ & $\rho=0.1$\\
$p=1.0$ & $p=0.01$\\
$v^x=0.99$ & $v^x=0$\\
$v^y=0$ & $v^y=0$\\\hline
bottom-left & bottom-right\\
$\rho=0.5$ & $\rho=0.1$\\
$p=1.0$ & $p=1.0$\\
$v^x=0$ & $v^x=0$\\
$v^y=0$ & $v^y=0.99$\\\hline
\end{tabular}
\caption{\label{tab:2dshock} Initial data for the 2D shock.}
\end{table}

\begin{figure}
\includegraphics[width=8.0cm]{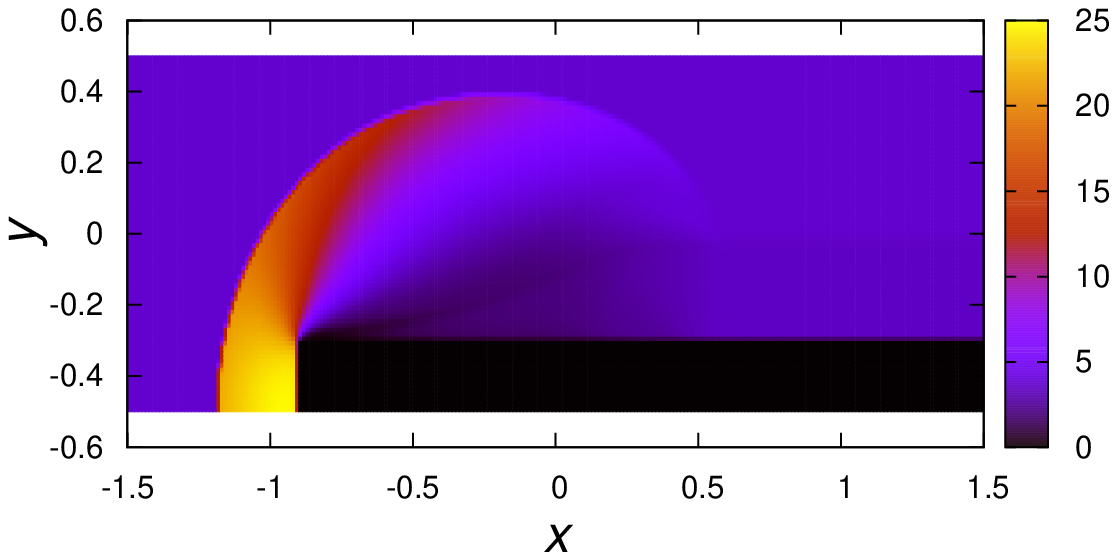}
\includegraphics[width=8.0cm]{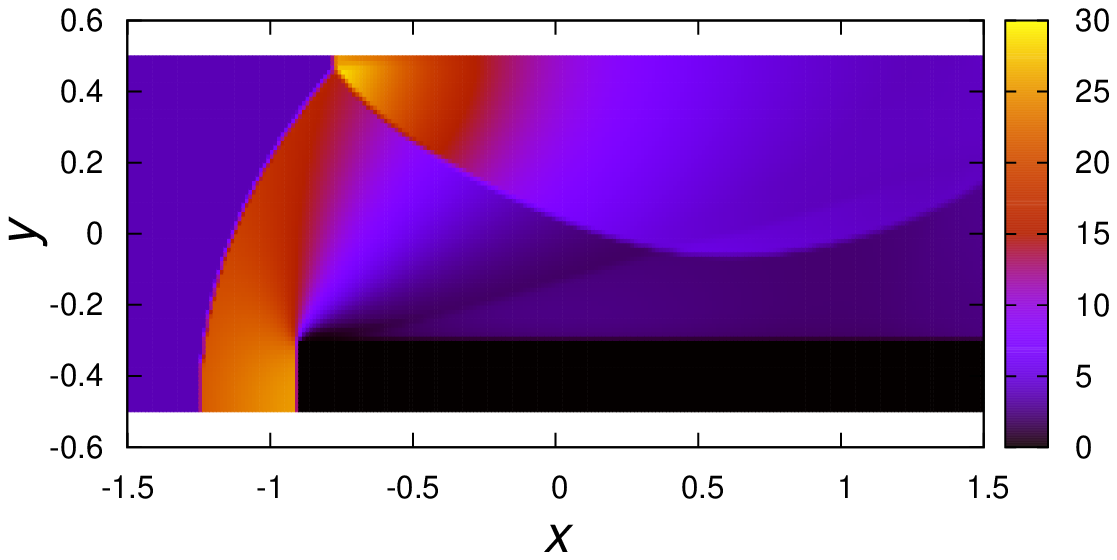}
\includegraphics[width=8.0cm]{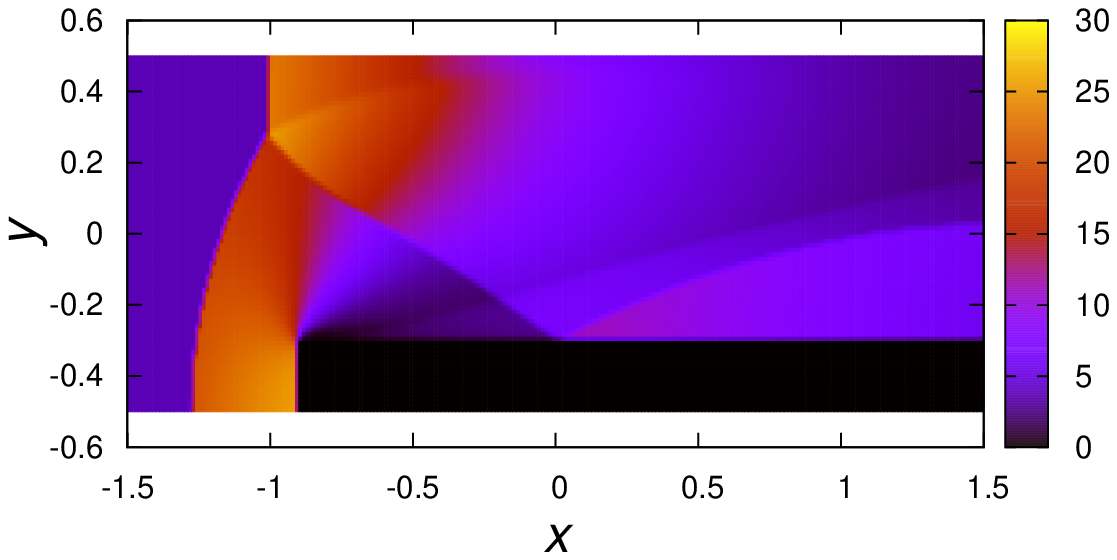}
\caption{\label{fig:tunel} Snapshots of the logarithm of the rest mass density for the Emery wind tunnel at times $t=1.56,~3.125,~4.69$. The step size is restricted to the domain $[-0.9,1.5]\times [-0.5,-0.3]$. The number of cells used is 1200$\times$400. It can be appreciated the bow shock when it forms an grows (first panel), when it bounces from the upper boundary (second panel) and when it bounces again from the step (third panel).}
\end{figure}

An additional test we want to include is the relativistic Emery's flat faced step. It consists in the flow entering from the left side of the domain and faces a step. The initial conditions are $\rho = 1.4$, $p = 1.0$, $v^x = 0.99$, $v^y = 0$ with $\Gamma = 1.4$. Three snapshots are shown in Fig. \ref{fig:tunel}. The boundary conditions are inflow at the left boundary, outflow at the right, reflection at the top and bottom and the step boundaries. Again we show only the morphology as in \cite{DonatFont}.

\begin{figure*}
\includegraphics[width=7.5cm]{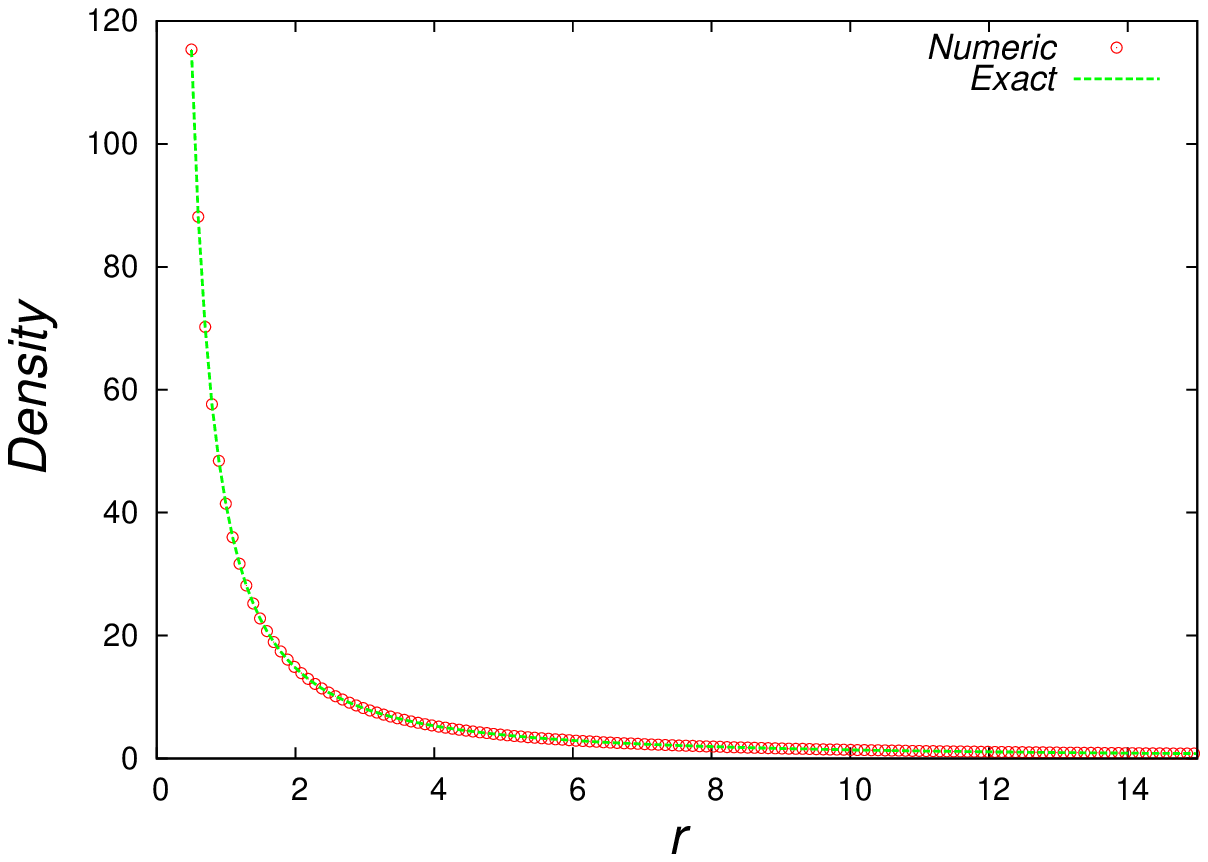}
\includegraphics[width=7.5cm]{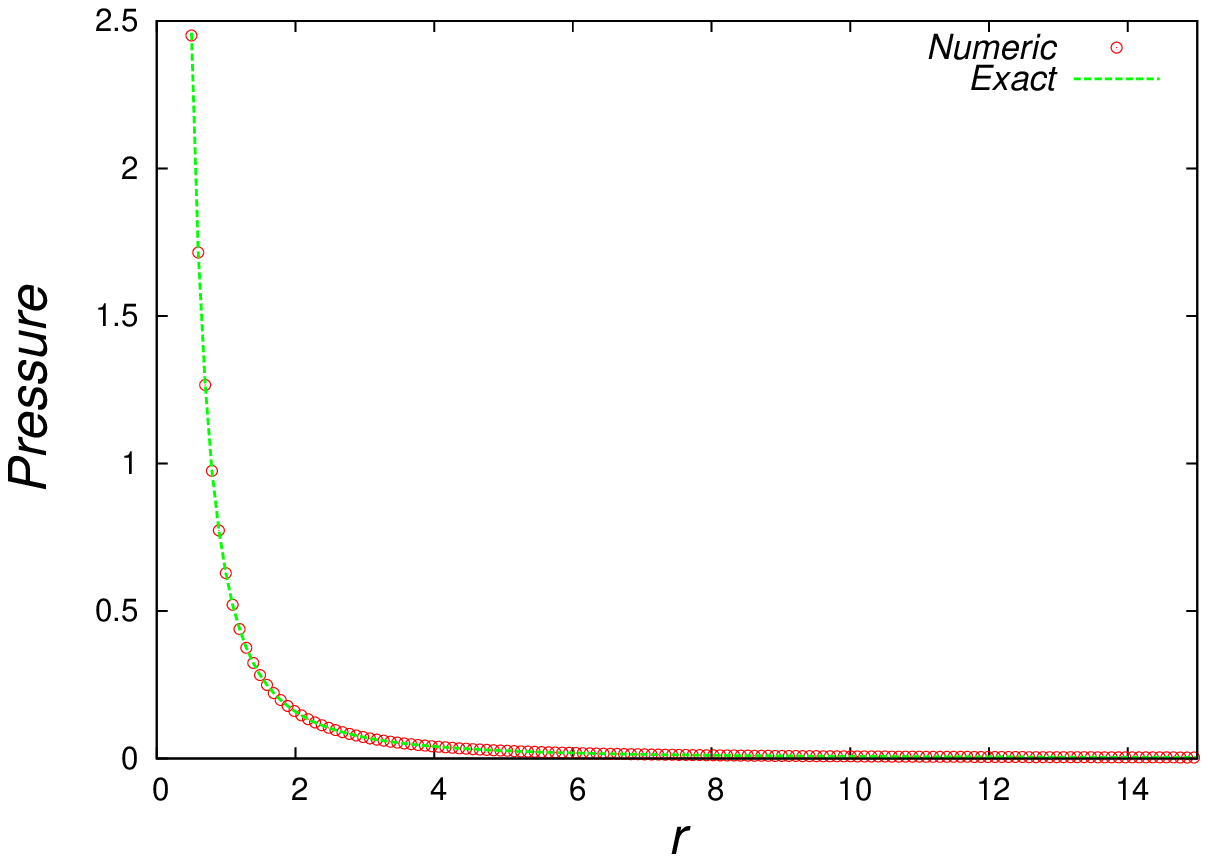}
\includegraphics[width=7.5cm]{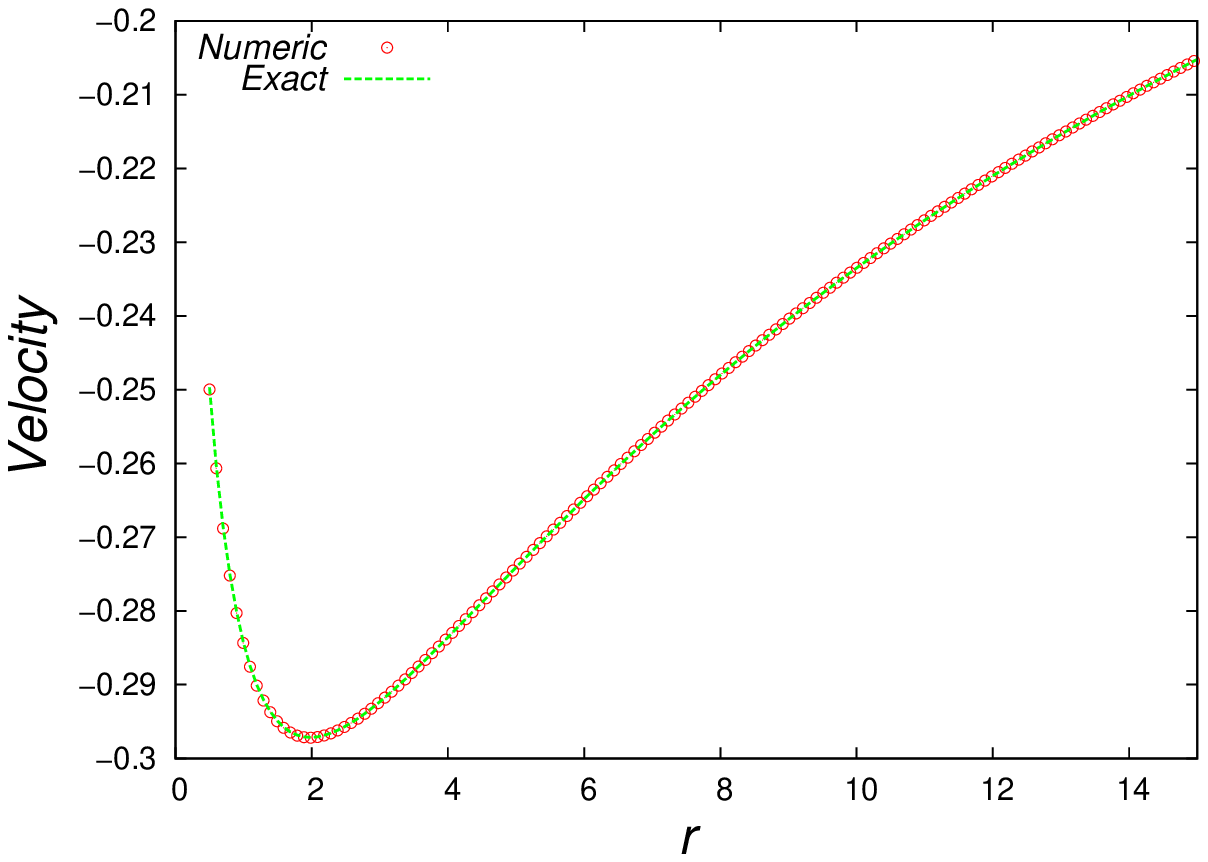}
\includegraphics[width=7.5cm]{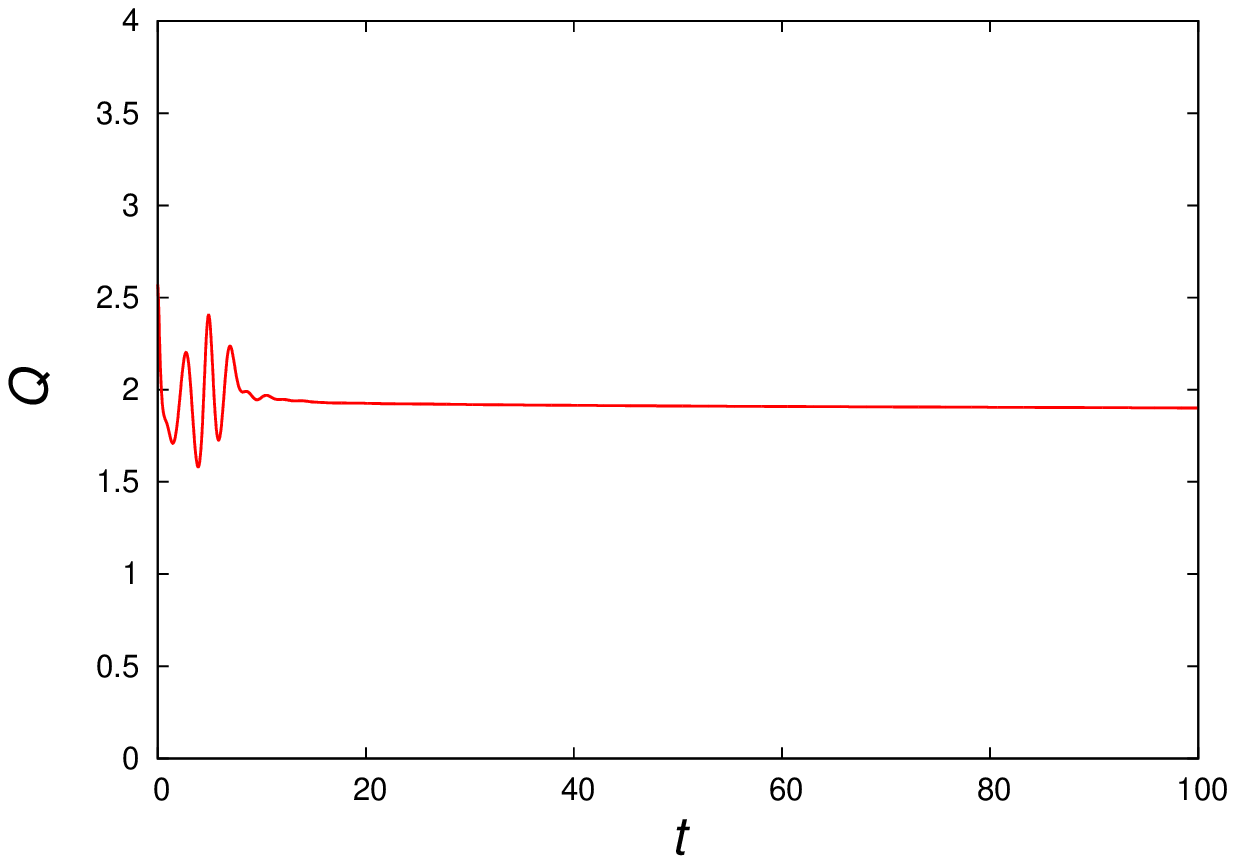}
\caption{\label{fig:michel} Numerical and exact solution of the Michel accretion at time $t=0.990M$. The solid line is the exact solution and the dots are the numerical solution. The last plot shows the convergence order calculated using 100,200,400 cells; this shows the nearly second order convergence of our implementation of this test. The time is in units of the black hole mass $M$.}
\end{figure*}

{\it Test 3, Michel accretion onto a black hole.} The standard test on curved space-times is the radial accretion of an ideal gas onto a Schwazschild black hole \cite{Michel1972}. For this, we constructed the exact solution in Eddington-Finkelstein coordinates using the parameters in \cite{Michel} at coordinate time $t=100M$. The results are shown in Fig. \ref{fig:michel}, where the distribution of the gas variables remains time-independent as expected. We also include a plot of the $L_2$ norm of the error in time for various resolutions, showing the convergence of the application.

\section*{Acknowledgments}

This research is partly supported by grants: 
CIC-UMSNH-4.9 and 
CONACyT 106466.
(FDLC) acknowledge support from the CONACyT.
The runs were carried out in NinaMyers computer farm.


\bsp

\label{lastpage}

\end{document}